\documentclass[11pt]{article}
\usepackage[utf8]{inputenc}
\usepackage[a4paper, margin=2.5cm]{geometry}
\usepackage{graphicx}
\usepackage{xcolor}
 \usepackage{float} 
\usepackage{latexsym}
\usepackage{comment}
\usepackage{multicol}
\usepackage{dsfont}
\usepackage{braket}
\usepackage{newunicodechar}
\newunicodechar{⁻}{$^{-}$}
\usepackage{amssymb,amsthm, amsmath}
\usepackage{hyperref}
\usepackage{subcaption}
\usepackage{caption}
\usepackage{mathtools} 
\usepackage{dsfont}
\usepackage{booktabs}
\usepackage{array}
\usepackage[backend=biber,sorting=none,giveninits=true,maxnames=99,maxcitenames=99]{biblatex}
\usepackage{soul}
\usepackage{authblk}
\usepackage{enumitem}
\usepackage{makecell}

\usepackage{hyperref}

\title{Resource-efficient variational quantum solver \\ for the travelling salesman problem \\ and its silicon photonics implementation}

\author[1,*]{Alessio Baldazzi}

\author[1]{Stefano Azzini}

\author[1]{Lorenzo Pavesi}

\affil[1]{ Department of Physics, University of Trento,Via Sommarive 14, Trento, 38123, Italy}

\affil[*]{ alessio.baldazzi@unitn.it }

\date{}

\begin{document}

\maketitle

\begin{abstract}
The travelling salesman problem is a well-known example of computationally-hard combinatorial problem for classical machines. 
Here, we propose a novel variational quantum algorithm to solve it. The method is based on the preparation of two maximally entangled quantum registers whose correlations are assigned to different paths between pairs of cities. For $N$ cities, this encoding requires $2 \lceil\log_2 N\rceil$ qubits and the solution to the problem is directly found in the correlation matrix of the two registers composing the overall trial state.
As a proof-of-concept experiment, we implement this algorithm for generic problems with four cities on a reconfigurable room-temperature silicon photonic circuit with integrated photon-pair sources, used to initialize maximally entangled path-encoded single-photon states.
\end{abstract}




\section{Introduction}
\label{sec:intro}

In classical combinatorial optimization, a paradigmatic problem with applications ranging from route finding to plotting and drilling machines is the Travelling Salesman Problem (TSP)~\cite{grotschel1991solution,grotschel1991optimal}. The TSP is easy to state but hard to solve~\cite{dantzig1954solution,beardwood1959shortest,bellmore1968tsp,grotschel1979traveling,reinelt1994traveling,cook2012pursuit,hoffman2013traveling}. There are $N$ cities, whose pairwise distances are known and collected in the $N\times N$ matrix $\emph{\textbf{D}}$. The goal consists in finding the shortest route visiting all cities once and returning to the starting city. The requirement of returning to the starting city does not alter the complexity of the problem.
The problem can be visualized as a directed weighted graph~\cite{lawler1985traveling,cook2012pursuit}. Without loss of generality, the graph can be taken fully connected since increasing sufficiently specific weights is equivalent to cutting the associated paths. If the matrix $\emph{\textbf{D}}$ is symmetric, then the graph is undirected, while in general the distances in two opposite directions can be different.
The TSP is one of the most studied problems in the field of combinatorial optimization and computational complexity. Its decisional formulation is NP-complete~\cite{karp1972reducibility,garey1979computers,lawler1985traveling,papadimitriou1994computational}: up to now, no classical algorithm can efficiently solve it with polynomial resources.

Let us dig on the problem to find a solver for this NP-complete problem.
A brute-force algorithm for an exact solution is impractical even for 10-20 cities, since the number of possible solutions is $(N-1)!$. The situation can be improved with the Held–Karp algorithm~\cite{held1962dynamic} using dynamic programming: it scales as $O(N^2\,2^N)$ in time and $O(N\,2^N)$ in space (memory), so it turns out to be infeasible even for 40 cities.
Other proposed approaches are optimization algorithms based on branch-and-bound, branch-and-cut and cutting planes~\cite{padberg1991branch,applegate2006traveling}, geometric algorithms with combinatorial optimization~\cite{grotschel1988geometric}, genetic algorithms~\cite{nagata1997powerful}, and heuristic methods~\cite{gutin2002traveling,karapetyan2011lin}. Besides the previous approaches, we also mention meta-heuristic algorithms~\cite{toaza2023review}, and Machine-Learning-based methods~\cite{alanzi2025tspml}. Generically, heuristic approaches, such as Nearest Neighbour and Lin–Kernighan search algorithms, are characterized by polynomial cost in terms of $N$, but they do not guarantee the optimality of the solution. However, because of the reduced resource cost and the good quality of the solutions, different heuristic and meta-heuristic methods, combined together, are mostly used nowadays, and they can solve instances with tens of thousand of cities with high precision.
On the other hand, the record for a classical solver using exact methods is 85.900 cities: it has been realized in almost a month through a cluster of 256 processors~\cite{applegate2006traveling}. 
Due to the classical computational intractability for exact solutions, TSP has also been investigated in the context of quantum computing (QC) and quantum simulation~\cite{Feynman82,divincenzo_physical_2000,bennett2000quantum,nielsen_chuang_2010,johnson2014quantum,Montanaro_2016}.
In particular, using the resources available today in the noisy intermediate-scale quantum (NISQ) era~\cite{Preskill_2018}, Variational Quantum Algorithms (VQAs)~\cite{Peruzzo_2014,cerezo,moll2017quantum,Shaydulin2021} and Quantum Approximate Optimization Algorithms (QAOAs)~\cite{FarhiGoldstoneGutmann2014,VerdonEtAl2019} have been used to formulate the TSP in the context of hybrid quantum-classical methods. 
More in general, combinatorial optimization problems - critical in industries like transportation, finance, logistics, and retail - remain challenging for both classical and quantum solvers~\cite{chen2025benchmarking}. 
This calls for further research to be carried out to understand the actual potential of NISQ devices for non-universal specific-purpose tasks.
Following this direction, the TSP represents a crucial benchmark.

\begin{figure}[t]
    \centering
    \includegraphics[width=\textwidth]{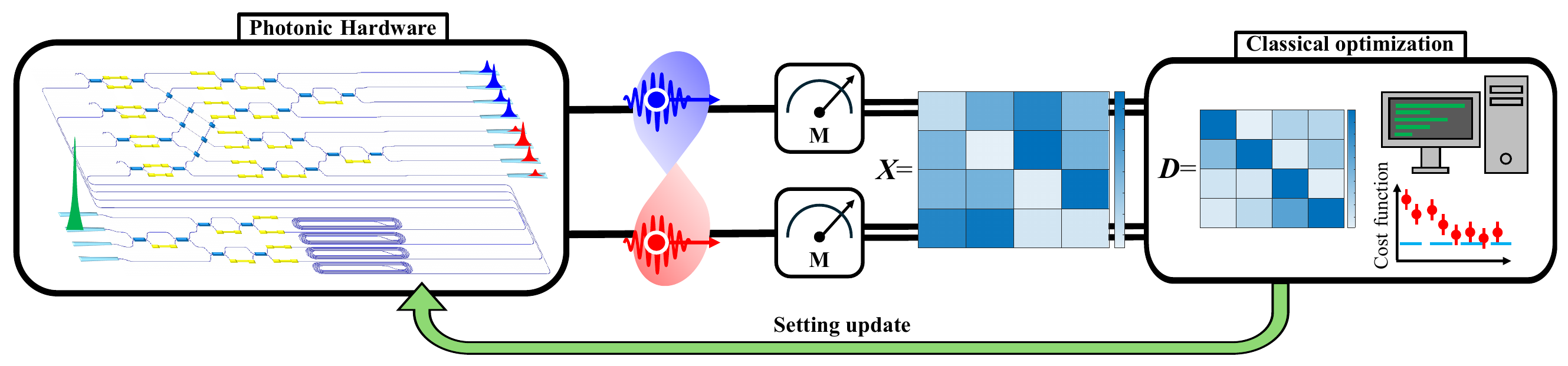}
    \caption{\textbf{Schematic workflow of the new VQA for TSPs.} 
    Like any VQA, the algorithm involves a reconfigurable quantum Hardware, on the left, and a PC, on the right. The first is able to prepare quantum trial states, and in our case it is based on a silicon photonic integrated circuit~\cite{Baldazzi2025}.
    The generic trial state is a bipartite maximally entangled system and its measurement produces the correlation matrix $\emph{\textbf{X}}$.
    In our photonic implementation, the system is composed of a pair of correlated photons generated on-chip.
    The PC evaluates the cost function given the correlation matrix $\emph{\textbf{X}}$ and the distance matrix $\emph{\textbf{D}}$ of the specific TSP, and then it executes an optimization routine which updates the setting used to prepare the trial state. 
    The variational algorithm ends when the cost function converges to its extremal value and the Hardware is trained to prepare the associated quantum state. }
    \label{fig:algorithm}
\end{figure}

In this work, we address this challenge and demonstrate that the TSP can be solved using a novel VQA. 
Crucially, the resulting quantum algorithm is particularly efficient in terms of the qubit number, scaling logarithmically with the number of cities. This feature makes it suitable for implementations on state-of-the-art photonic integrated circuits (PICs).
Indeed, silicon-on-insulator PICs~\cite{Wang_2018,adcock2019programmable,wang2019boson,Llewellyn_2019,Bartlett2020Universal,wang2020integrated,Vigliar_2021,Adcock_2021,Lee_24,Baldazzi2025} represent an already validated platform to implement VQAs, due to its well-established technology able to linearly manipulate photon states with e.g. Mach-Zehnder interferometer (MZI), and to create entangled photon states with e.g. parametric photon pair sources~\cite{harris_2016,qiang2018large,bogaerts2018silicon,bogaerts2020programmable,moody2022,luo2023recent,Yu_24,huang2024demonstration}. 
In~\cite{Baldazzi2025}, we show the capabilities of silicon photonics VQAs through a room-temperature silicon photonic integrated circuit (Si-PIC) implementing a four-qubit variational quantum eigensolver able to solve a quantum chemistry problem and factorization tasks with high accuracy. 
Here, we solve TSPs with four cities through the same small-scale and simple photonic processor with integrated sources as in~\cite{Baldazzi2025}. 
Figure~\ref{fig:algorithm} shows the sketch of our implemented approach to TSPs. 
Our VQA exploits an integrated reconfigurable photonic hardware, on the left, to prepare trial states made of a maximally entangled bipartite state of two photons. The correlation matrix $\emph{\textbf{X}}$ of the trial state is evaluated and a PC computes the cost function associated with the desired TSP, encoded in the distance matrix $\emph{\textbf{D}}$. Through a classical optimization routine, the trial state is updated. The iteration ends with the convergence of the cost function.

The manuscript is organized as follows.
Section~\ref{sec:class_form} contains a review of the classical formulation of the TSP in order to specify the problem and introduce the definitions used in the rest of the paper. In section~\ref{sec:literature} we briefly examine the literature of quantum approaches to the TSP. Section~\ref{sec:var_algorithm} describes the new VQA to solve TSPs, and Section~\ref{sec:experiments} shows the results of the proposed VQA for some configuration of four cities obtained with our Si-PIC.
Finally, in Section~\ref{sec:disc}, we comment on the scalability of the method and compare it with other quantum approaches to the TSP.
Further details are reported in the Appendices~\ref{app:prop_X} and~\ref{app:4city}.


\section{Classical formulation for TSP}
\label{sec:class_form}

\begin{figure}[t]
    \centering
    \includegraphics[scale=0.35]{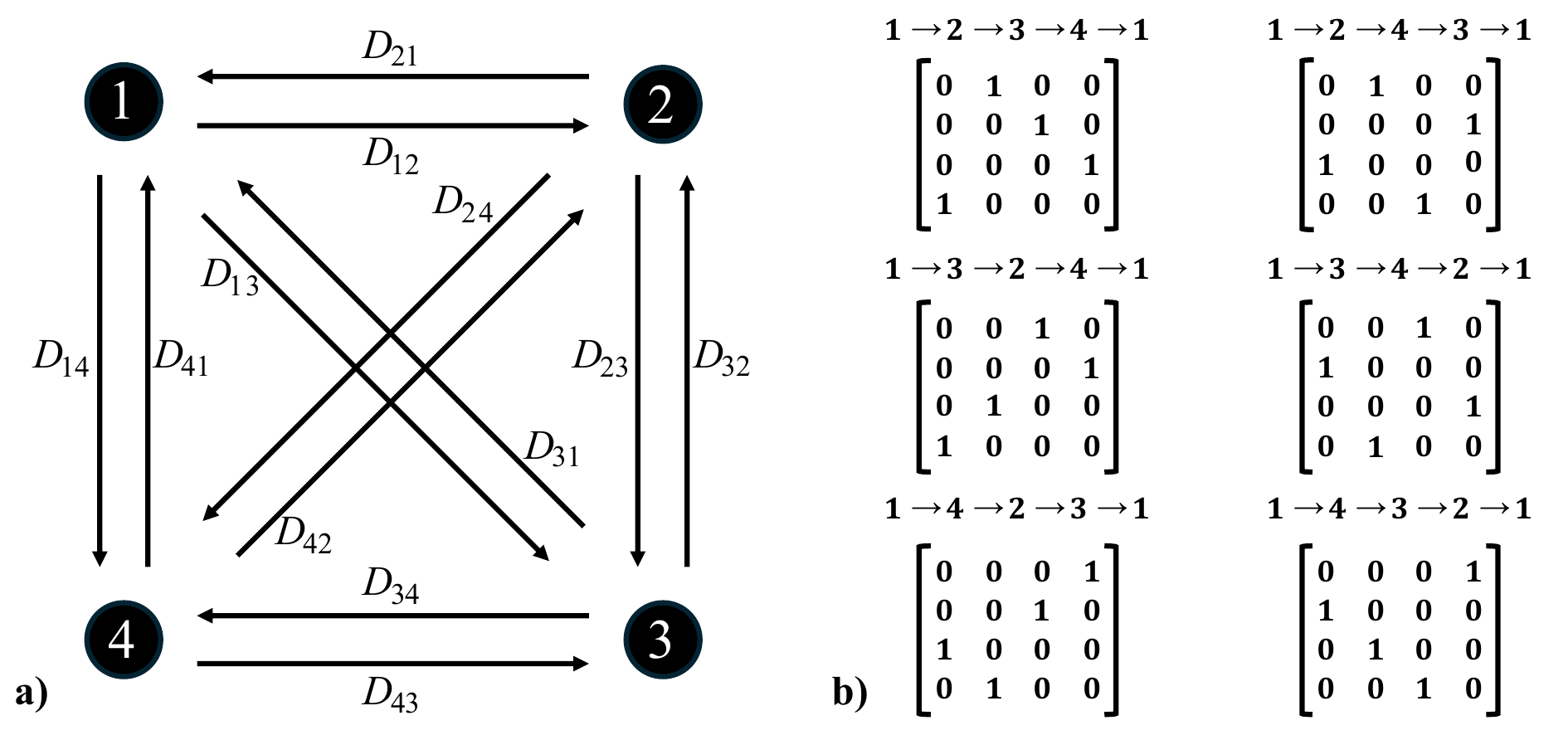}
    \caption{
    \textbf{Travelling salesman problem with four cities.}
    \textbf{(a)} Fully connected directed weighted graph associated with the travelling salesman problem with four cities. The parameters $D_{ij}$ quantify the distance of the path departing from city $i$ and arriving at city $j$. 
    Paths with the same departing and arriving city, associated with diagonal terms of the distance matrix $\emph{\textbf{D}}$, have not been represented: these terms can be chosen large enough in such a way to avoid solutions not satisfying the constraint $\textbf{(0)}$ in Eq.~\eqref{eq:class_constraints}.
    \textbf{(b)} Route adjacency matrices $\emph{\textbf{x}}$, defined in Eq.~\eqref{eq:class_bin_var}, representing all the routes among four cities satisfying all the constraints in Eq.~\eqref{eq:class_constraints}. The associated route is reported above each matrix. Starting and ending with city 1 is just a choice: any cyclic permutation of the route does not alter the associated $\emph{\textbf{x}}$ and the route length, because of the cyclicity of the problem.
    }
    \label{fig:class_form}
\end{figure}

The problem can be expressed with the following integer linear programming formulation~\cite{nemhauser1988integer}. Let us consider binary variables arranged in  the $N\times N$ matrix $\emph{\textbf{x}}$, called route adjacency matrix, whose components are defined as follows
\begin{equation}
    x_{ij} =
    \begin{cases}
        1 \quad, \, \mbox{if the path departs from city $i$ and arrives at city $j$;} \\
        0 \quad, \, \mbox{otherwise,} 
    \end{cases}
    \label{eq:class_bin_var}
\end{equation}
with $i$ and $j$ in $[1,N]$. The goal is to minimize the route length, i.e.
\begin{equation}
    \sum_{i=1}^N \sum_{j=1}^N D_{ij}\,x_{ij} \,,
    \label{eq:class_route_length}
\end{equation}
where $D_{ij}$, component of $\emph{\textbf{D}}$, is the distance from city $i$ to city $j$, and $N$ the number of cities. 
Without any constraint, the minimization of the route length will generally produce unacceptable routes. To prevent this, the binary variables $x_{ij}$ must satisfy the following constraints:
\begin{equation}
\begin{split}
    \textbf{(0)} & \quad x_{ii} = 0 \hspace{1.5cm},\quad \forall\,i\in[1,N] \,,
    \\
    \textbf{(1)} & \quad \sum_{i=1}^N x_{ij} = 1 \qquad,\quad \forall\,j\in[1,N] \,,
    \\
    \textbf{(2)} & \quad \sum_{j=1}^N x_{ij} = 1 \qquad,\quad \forall\,i\in[1,N] \,,
    \\
    \textbf{(3)} & \quad \sum_{i\in S}
    \sum_{j\notin S} x_{ij} \ge 1 \hspace{0.55cm},\quad \forall\,S \subseteq \{1\ldots N\}\,,\,0 < |S| < N \,.
\end{split}
\label{eq:class_constraints}
\end{equation}
The constraint $\textbf{(0)}$ eliminates paths with the same departing and arriving city.
This constraint can be removed by setting sufficiently large values in the diagonal components of $\emph{\textbf{D}}$. This implies that paths with the same departing and arriving city are "dynamically" excluded during the optimization process. 
The constraint $\textbf{(1)}$/$\textbf{(2)}$ means that each city is the arrival from/departure to exactly one city. These two sets of constraints make $\emph{\textbf{x}}$ a doubly-stochastic matrix~\cite{birkhoff1946,horn2012matrix}, and, since it is a binary matrix, consequently a permutation matrix~\cite{brualdi1991combinatorial}. Constraints $\textbf{(1)}$-$\textbf{(2)}$ imply also that the sum of all the components of $\emph{\textbf{x}}$ is equal to $N$. Finally, the constraint $\textbf{(3)}$ is denoted as the subtour elimination constraint~\cite{dantzig1954solution,nemhauser1988integer}. The latter relations ensure that the found solution is composed of a single route and not of a union of disconnected routes. The constraint $\textbf{(3)}$ can be expressed in other equivalent ways, but the one here used is typically more efficient. For example, in the Miller–Tucker–Zemlin formulation~\cite{miller1960integer} dummy variables, keeping track of the visit order, are introduced to eliminate solutions with subtours. This formulation requires large time instances because of the model enlargement. Figure~\ref{fig:class_form} shows the generic graph and allowed matrices $\emph{\textbf{x}}$ with associated routes for the general TSP with four cities.

Note that the constraints $\textbf{(1)}$-$\textbf{(2)}$ restrict the rows or columns of $\emph{\textbf{x}}$ to be a set of orthonormal vectors in $\mathbb{R}^N$. 
Using this property, we find that solutions allowed by these constraints can be written as follows
\begin{equation}
    \emph{\textbf{ x }} = \sum_{k = 1}^N \textbf{e}_k \cdot \textbf{e}_{\sigma(k)}^{\rm T} 
    = \sum_{k = 1}^N \textbf{e}_k \otimes \textbf{e}_{\sigma(k)} 
    = \sum_{k = 1}^N \textbf{e}_k \otimes \left(\textbf{P}_{\sigma} \cdot \textbf{e}_k\right) \,,
    \label{eq:classicalBV_super}
\end{equation}
where ${\rm T}$ denotes the transpose, $\{ \textbf{e}_k\}_{k \in (1\ldots N)}$ is the canonical basis of $\mathbb{R}^N$, $\sigma$ is a permutation such that $\sigma(k)\ne k$, $\forall k \in (1\ldots N)$, and $\textbf{P}_{\sigma}$ is the associated rotation. Such a discrete map is permuting the elements of the canonical basis, and $\{\textbf{P}_{\sigma} \cdot \textbf{e}_k\}_{k \in (1\ldots N)}$ is a permutation of the canonical basis. In this form, $\emph{\textbf{x}}$ is a linear combination of outer products of two vectors, one representing the departure and one the arrival of each path of a specific route.
Note also that in this form no temporal order is assigned to the different cities. We could rearrange the order of the terms in the previous sum without changing the route, and consequently its length. Indeed, the TSP is cyclic and the route is defined modulo the starting city's assignment.

We conclude this section by stressing that the matrix $\emph{\textbf{x}}$ contains all the information about a specific route. 
Thus, once the optimal $\emph{\textbf{x}}$ is found, the TSP is solved.

\section{Quantum approaches to TSP}
\label{sec:literature}

The TSP with $N$ cities can be formulated as a quadratic unconstrained binary optimization (QUBO) problem in terms of Ising-like Hamiltonians by introducing $N^2$ binary variables and obtaining a quadratic Hamiltonian containing penalty terms for not-allowed routes~\cite{Lucas_2014,egger2021quantum}. In this case, the dimensionality of the associated Hilbert space is $2^{N^2}$.
Starting from this formulation, $N$ quantum systems with $N$ levels are considered~\cite{vargas2021manyqudit}: the use of qudits~\cite{Wang_2020,meth2025simulating} reduces the dimension of the Hilbert space of the Ising formulation from $2^{N^2}$ to $2^{N \log_2 N}$. Converting the previous formulation into the qubit language, it is possible to reduce the number of qubits from $N^2$ to $N \log_2N$ for QAOAs~\cite{ramezani2024reducing}.
This reduction has a trade-off with respect to the number of two-qubit gates: they scale as $O(N^3)$ for $N^2$-encoding, and as $O(N^4 \log_2 N)$ for $(N \log_2N)$-encoding.
An alternative formulation of QAOA to solve the TSP is given in~\cite{ruan2020quantum}, where each edge is mapped to one qubit state, and state one means that the edge belongs to the route, otherwise it does not. In this way, the quantum register is made of $N(N-1)/2$ qubits, and the constraints are introduced in the Hamiltonian mixer term~\cite{HenSpedalieri2016}.
Different space-efficient encodings have been considered, such as ranking and Lehmer codes, for different QAOAs~\cite{Glos2022SpaceEfficient,BourreauFleuryLacomme2023}. 
In addition, quantum walk techniques~\cite{ChildsGoldstone2004,MarshWang2019}, VQE with Heisenberg exchange gates~\cite{ellertbeck2021magnetism}, quantum phase estimation algorithm~\cite{kitaev1995,Cleve_1998,nielsen_chuang_2010} with distances encoded into phases of unitary operators~\cite{srinivasan2018efficient}, 
variational quantum circuits equipped with quantum self-attentive neural networks~\cite{ruan2024quantum}, 
a quantum heuristic Grover-like algorithm~\cite{BangYooLimRyuLee2012},
quantum dynamic programming to generate a superposition of TSP solutions~\cite{xujun2025quantumspeedupalgorithmtsp}, hybrid quantum-classical approaches through quantum optimisation techniques with classical machine learning methods~\cite{lytrosyngounis2025} are proposed to solve the TSP. Besides these approaches, we also mention a theoretical proposal and simulations based on a single qubit~\cite{goswami2024solving,Baniata_25}. In this method, each city is represented by a point on the Bloch sphere, the distances are encoded in scalar products between the city states and auxiliary city-states, and parametrized single-qubit rotations allow to create superposition of different routes. 
The TSP with the Ising formulation has also been studied by using Adiabatic Quantum Optimization\cite{HeimBrownWeckerTroyer2017}, Quantum Annealing~\cite{KadowakiNishimori2002,MartonakSantoroTosatti2004}, Quantum Annealer D-Wave~\cite{mcgeoch2013experimental,Jain2021_TSP_DWave,stogiannos2022experimental}, Quantum Annealing together with neural network~\cite{He2024_QA_GNN_TSP}, time windows with QUBO and HUBO formulations on quantum annealing devices~\cite{SalehiGlosMiszczak2021}. 
Finally, for completeness, we mention analyses of different QAOAs with gate-based digital quantum simulators and associated mixer designs in terms of precision, computational cost and robustness against noise in different configurations of 3 to 5 cities~\cite{qian2023comparative}, the study on the dependence with respect to various hyperparameters, such as the classical optimizer choice and strength of the TSP constraint penalization for QAOAs and VQAs~\cite{PalackalPoggelWulff2023}, 
the comparison of different approaches, considering Simulated Annealing, QUBO methods on quantum annealers and Optical Coherent Ising Machines, different QAOAs and VQAs together with different encodings, and the Quantum Phase Estimation algorithm on gate-based quantum computers and simulators~\cite{zhu2022realizablegasbasedquantumalgorithm,schnaus2024efficient,PadmasolaEtAl2025,smith2025travelling,Sato_2025}.

\section{New variational quantum algorithm to solve TSP}
\label{sec:var_algorithm}

Here, we present a method based on the generalization of the classical route adjacency matrix $\emph{\textbf{x}}$ to its quantum version $\emph{\textbf{X}}$, whose properties are described in Appendix~\ref{app:prop_X}.

\begin{figure}[t]
    \centering
    \includegraphics[scale=0.45]{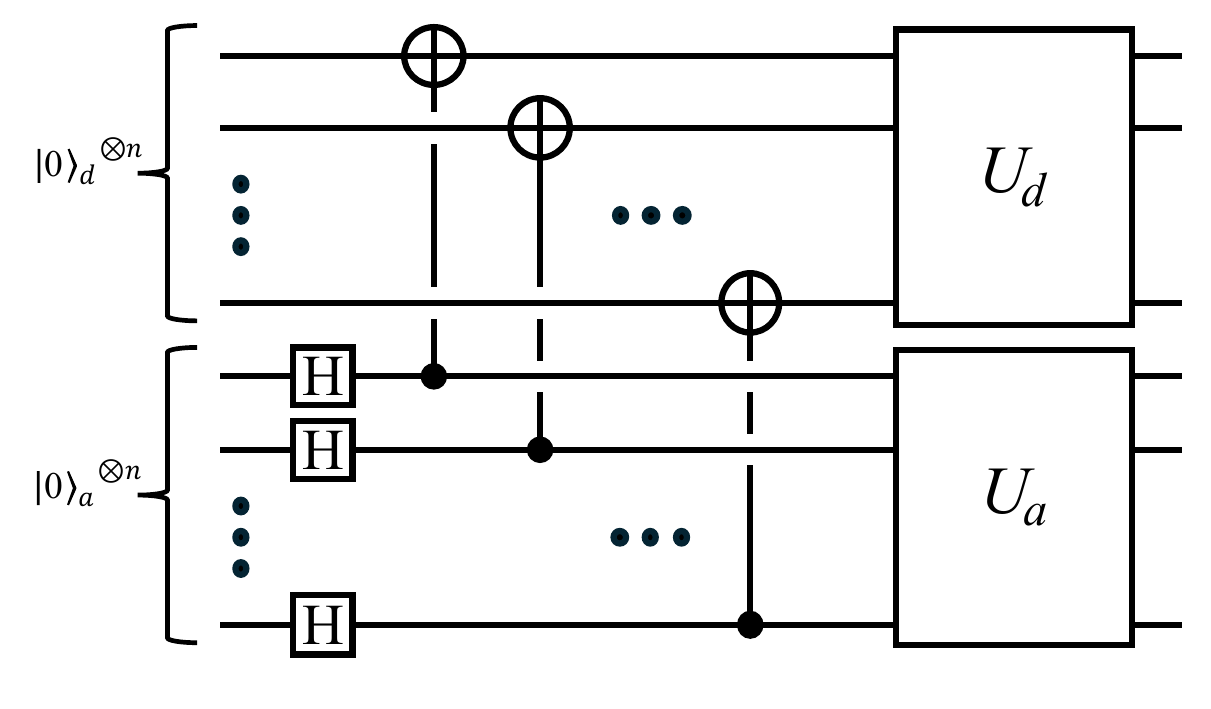}
    \caption{
    \textbf{Quantum circuit to prepare the trial states for our VQA for TSPs.}
    Gate representation of the quantum circuit able to prepare the generic trial state for the presented VQA associated with TSPs for $N$ cities. First of all, $2n = 2 \lceil\log_2 N\rceil$ qubits are collected in two registers labelled with $d$ for "departures" and $a$ for "arrivals", and they are initialized to the zero state. Then, Hadamard gates (H) are applied to each qubit in the arrival register, and controlled-NOT gates are applied pairwise to couple of qubits, one in the departure and one in the arrival register, taking the second as the control. At this point, the two registers form a maximally entangled state, Eq.~\eqref{eq:max_ent}. Finally, two independent unitary transformations $U_{d/a}$ are applied to the departure and arrival registers.
    The final state, reported in Eq.~\eqref{eq:final_prep_state}, can be used as a trial state for the presented VQA associated with the cost function in Eq.~\eqref{eq:cost_funct}.
    }
    \label{fig:quantum_circuit}
\end{figure}

Let us take two registers, each containing $n\equiv\lceil\log_2 N\rceil$ qubits:
one for departures $\{| k \rangle_d \}_{k \in\{0,1\}^n}$ and one for arrivals $\{| k \rangle_a \}_{k \in\{0,1\}^n}$.
Starting from the state $| 0 \rangle_d^{\otimes n} \otimes | 0 \rangle_a^{\otimes n}$, we entangle the two registers pairwise, obtaining the following state
\begin{equation}
    |\psi_0\rangle = \frac{1}{2^{n/2}}\sum_{k \in\{0,1\}^n} | k \rangle_d \otimes | k \rangle_a \,.
    \label{eq:max_ent}
\end{equation}
Then, we apply a unitary transformation, i.e. $U_{d/a}$, to each register.
The associated quantum circuit is shown in Figure~\ref{fig:quantum_circuit}. The final state reads as follows
\begin{equation}
    |\psi\rangle = \frac{1}{2^{n/2}}\sum_{k \in\{0,1\}^n} | \xi_k^{(d)} \rangle_d \otimes | \xi_k^{(a)} \rangle_a \,,
    \label{eq:final_prep_state}
\end{equation}
where $|\xi_k^{(d/a)}\rangle \equiv U_{d/a} | k \rangle_{d/a}$ , with $k \in \{0,1\}^n$.
We decide to transform both registers, but it is possible to transform just one register without loss of generality because of the cyclicity of the problem and the properties of maximally entangled states, e.g. $\left[U_{d}\otimes U_{a}\right] | \psi_0 \rangle = \left[\left(U_{d}\cdot U_{a}^{\rm T}\right)\otimes \mathbf{1} \right]| \psi_0 \rangle$~\cite{khatri2019quantum,xue2022variational}. The transformation of both registers can be preferred if used together with some constraints on $U_{d}\otimes U_{a}$. Appendix~\ref{app:4city} contains an example of these transformations for the generic TSP with four cities.\\
The bipartite state~\eqref{eq:final_prep_state} is a maximally entangled state of the two registers; indeed, a maximally mixed state is obtained by taking the partial trace of one register. Note also that the state~\eqref{eq:final_prep_state} can be mapped to the Hilbert space associated with two qudits, each with dimension $N$.\\
If $N$ is not a power of two, the states $\{|k\,(\mbox{mod}\,2) \rangle_{d}\otimes |k\,(\mbox{mod}\,2) \rangle_{a}\}_{k\in\mathcal{B}}$ with $\mathcal{B}=(N,2^n]$ in Eq.~\eqref{eq:final_prep_state} do not represent any city of the problem. Thus, we set the transformation $U_d\otimes U_a$ to have a trivial action on those states, which will be spectator qubits.\\
To simplify the notation, we drop the subscripts $d/a$ in the two registers' states for the rest of the manuscript: reading from the left, the first component of the total state is the departure register and the second the arrival register.

At this point, by labeling cities from $1$ to $N$, we make the following crucial assignment:
\begin{equation}
\begin{split}
    \mbox{the path departing from ci} &\mbox{ty $i$ and arriving at city $j$} \\
    &\updownarrow  \\
    | i-1\,(\mbox{mod}\,2) \rangle_d &\otimes | j-1\,(\mbox{mod}\,2) \rangle_a \,.
    \label{eq:path_state}
\end{split}
\end{equation}
With respect to the computational basis, each term contained in state $|\psi\rangle$, Eq.~\eqref{eq:final_prep_state}, represents an edge. 
For a generic transformation $U_{d}\otimes U_a$, the total state $|\psi\rangle$ is a superposition of states representing different paths. Grouping different computational basis states is equivalent to creating different routes. (At this stage, the grouping is arbitrary, but the choice of the observable of our VQA is going to remove this arbitrariness.) Therefore, the total state $|\psi\rangle$ is also a superposition of states representing different routes, where some do not satisfy all the constraints in Eq.~\eqref{eq:class_constraints}, and others do.
The previous consideration implies that the final state $|\psi\rangle$ in Eq.~\eqref{eq:final_prep_state} can be used as the generic trial state, if such a state, or equivalently the transformation $U_{d}\otimes U_{a}$, has enough expressibility~\cite{Tangpanitanon_2020,Holmes_2022}. Indeed, in such a case, the final state $|\psi\rangle$ in Eq.~\eqref{eq:final_prep_state} has a non-zero overlap with the state associated with the optimal route.

At this point, we can observe similarities between the classical formulation and the trial state $|\psi\rangle$. 
First of all, since each register in state $|\psi_0\rangle$ contains all the elements of the canonical basis and $U_{d/a}$ are unitary maps, the set of different states present in each register of $|\psi\rangle$ is still an orthonormal basis of $\mathbb{C}^n$, i.e.
$
   \langle \xi_k^{(d/a)} | \xi_l^{(d/a)} \rangle
   =\delta_{kl} 
   \,,\,\, \forall\,k,l\in(1..N) \,.
$
From this perspective, Eq.~\eqref{eq:final_prep_state} resembles the form of the classical binary variable in Eq.~\eqref{eq:classicalBV_super}: in both cases, we have a linear combination of paths and the vectors of departures/arrivals form an orthonormal basis.
In particular, because of the assignment in Eq.~\eqref{eq:path_state}, the route described in Eq.~\eqref{eq:classicalBV_super} corresponds to the following state:
\begin{equation}
    |\psi_{\sigma}\rangle = \frac{1}{2^{n/2}}\sum_{k \in\{0,1\}^n} | k \rangle \otimes | \sigma(k) \rangle \,,
    \label{eq:route_quantum_state}
\end{equation}
where $\sigma$ is the same permutation ($\sigma(k)\ne k$, $\forall k \in (1\ldots N)$). Like the classical description, we have cyclicity, since the change of summation order is irrelevant. This means that different transformations $U_d\otimes U_a$ could give rise to the same route.
We anticipate that the states~\eqref{eq:route_quantum_state} are equivalent to states with the same computational basis terms and different relative phases, since the needed observables are evaluated on the computational basis. This redundancy can be partially removed by choosing the unitary transformation $U_d\otimes U_a$ to be real symmetric.

Based on the previous considerations, the specific-route information analogous to that stored in the classical route adjacency matrix $\emph{\textbf{x}}$ can be obtained by the measurement of the state~\eqref{eq:route_quantum_state} on the computational basis. Thus, the quantum version $\emph{\textbf{X}}$, the quantum route adjacency matrix, is defined as follows
\begin{equation}
\begin{split}
    X_{ij}(\boldsymbol{\alpha}) & \equiv 
    2^n \, \mbox{Tr}[ \rho(\boldsymbol{\alpha})\, \hat{P}_{ij} ] \,,
    \label{eq:quantum_BV}
\end{split}
\end{equation}
where $\rho = |\psi\rangle\langle\psi|$ is the density matrix, $\boldsymbol{\alpha}$ are the variational parameters characterizing the transformation $U_d\otimes U_a$, and $\hat{P}_{ij} = \left(|i \rangle \otimes |j \rangle\right) \,\left(\langle i | \otimes \langle j | \right)$, with $(i,j) \in \{0,1\}^n$, are the canonical von Neumann projectors.
Basically, $X_{ij}$ represents the correlation of the two registers with respect to the canonical basis element $|i \rangle \otimes |j \rangle$. \\
Note that $X_{ij} \ge 0$ due to the positivity of $\rho$, and
\begin{equation}
\begin{split}
    \sum_{i\in \{0,1\}^n} X_{ij} = 1
    \quad,\, \forall\,j\in\{0,1\}^n \,,
    \quad\mbox{and}\quad
    \sum_{j\in \{0,1\}^n} X_{ij} = 1
    \quad,\, \forall\,i\in\{0,1\}^n \,,
\end{split}
\end{equation}
which are the analogous of constraints $\textbf{(1)}$-$\textbf{(2)}$ in Eq.~\eqref{eq:class_constraints}. Appendix~\ref{app:prop_X} contains the proof of the $\emph{\textbf{X}}$'s properties. 
These identities ensure that departing from/arriving at all cities in the departure/arrival register has 100\% probability for each arrival/departure state. Therefore, the observable $\emph{\textbf{X}}$ is a doubly-stochastic matrix by construction, without imposing the constraints $\textbf{(1)}$-$\textbf{(2)}$. 
The classical case is recovered when there is 100\% probability of departing from/arriving at one city in the departure/arrival register for each city in the arrival/departure register. The correlation matrix of states shown in Eq.~\eqref{eq:route_quantum_state} is precisely a permutation matrix. In Appendix~\ref{app:prop_X}, we show that a correlation matrix equal to a permutation matrix corresponds to a state given in Eq.~\eqref{eq:route_quantum_state}, modulo relative phases between the computational basis terms.
In this case, $\emph{\textbf{x}}^* = \emph{\textbf{X}}(\boldsymbol{\alpha}^*)$, where $\emph{\textbf{x}}^*$ is a permutation matrix representing a route that satisfies constraints $\textbf{(1)}$-$\textbf{(2)}$ and $\boldsymbol{\alpha}^*$ is the associated setting of the variational parameters, which is generally not unique. This parametrization redundancy in the solution representation is due to the cyclic nature of the TSP.
Different transformations $U_d\otimes U_a$ could correspond to the same state, where the terms in the superposition are simply rearranged, as pointed out in Eq.~\eqref{eq:route_quantum_state}.
Moreover, relative phases among the different terms of the states in Eq. \eqref{eq:route_quantum_state} do not alter the correlation matrix $\emph{\textbf{X}}$ between the two registers. By choosing $U_d\otimes U_a$ to be a symmetric map, we reduce the number of free, or variational, parameters.\\
More generally, given the assignment in Eq.~\eqref{eq:path_state}, $\emph{\textbf{X}}$ is a linear combination of different classical $\emph{\textbf{x}}$ satisfying constraints $\textbf{(1)}$-$\textbf{(2)}$.
By using the Birkhoff–von Neumann theorem~\cite{birkhoff1946}, $\emph{\textbf{X}}$ is a convex combination of different $\emph{\textbf{x}}$s, since the generic $\emph{\textbf{X}}$ is a doubly-stochastic matrix and classical $\emph{\textbf{x}}$s permutation matrices. This means that
\begin{equation}
    \emph{\textbf{X}}(\boldsymbol{\alpha}) = \sum_k \lambda_k(\boldsymbol{\alpha}) \, \emph{\textbf{x}}_k
    \quad, \quad \sum_k \lambda_k =1 \,\,\mbox{and}\,\,\lambda_k\ge 0 \;\; \forall k\,,
    \label{eq:birkhoff}
\end{equation}
where $\{\emph{\textbf{x}}_k\}_k$ are permutation matrices.
This implies that the quantum route adjacency matrix $\emph{\textbf{X}}$ represents a convex combination of classical routes. Note that the grouping of the different terms discussed after Eq.~\eqref{eq:path_state} does not matter: $\emph{\textbf{X}}$ will always be a linear combination of correlation matrices associated with states reported in Eq.~\eqref{eq:route_quantum_state}, which correspond to classical routes. However, the Birkhoff–von Neumann decomposition is unique only when $\emph{\textbf{X}}$ is a permutation matrix~\cite{Dufosse2018}. Such a non-uniqueness is due to the geometry of the Birkhoff–von Neumann polytope (the space of doubly-stochastic matrices) and the presence of cycles in the support graph of the associated doubly-stochastic matrix. We anticipate that our variational algorithm is not affected by this ambiguity, since the cost function is a function of $\emph{\textbf{X}}$ itself, and it does not depend on the individual $\{\lambda_k\}_k$ of a specific decomposition.\\
If $N$ is not a power of two, since the transformation $U_d\otimes U_a$ has a trivial action on the states $\{|k\,(\mbox{mod}\,2) \rangle\otimes|k\,(\mbox{mod}\,2) \rangle\}_{k\in\mathcal{B}}$ with $\mathcal{B}=(N,2^n]$, $\emph{\textbf{X}}$ is block diagonal where the second block with indexes $(i,j)\in (N,2^n]\times (N,2^n]$ is the identity matrix. Equivalently, the permutation matrices $\{\emph{\textbf{x}}_k\}_k$ composing the total $\emph{\textbf{X}}$ correspond to permutations that leave invariant the states that do not represent any city.

The cost function is built by replacing the binary matrix $\emph{\textbf{x}}$ in the route length in Eq.~\eqref{eq:class_route_length} with the correlation matrix $\emph{\textbf{X}}$ and adding a term for subtour elimination:
\begin{equation}
    {\rm C}(\boldsymbol{\alpha}) = \sum_{i,j \in \{0,1\}^n} D_{ij}\,X_{ij}(\boldsymbol{\alpha})
    -A_{\rm sub} \!\!\! \sum_{\substack{S \subseteq \{1\ldots N\}\\0 < |S| < N} } \sum_{i \in S } \sum_{j \notin S } X_{ij}(\boldsymbol{\alpha}) \,,
    \label{eq:cost_funct}
\end{equation}
where the diagonal terms of $\emph{\textbf{D}}$ and the coefficient $A_{\rm sub}$ are chosen large enough to suppress diagonal terms $X_{ii}$ $\forall\,i\in\{0,1\}^n$, representing the paths with the same departing and arrival city, and solutions with subtours, respectively. In this way, the constraints $\textbf{(0)}$ and $\textbf{(3)}$ of Eq.~\eqref{eq:class_constraints} are implemented during the optimization process. In particular, non zero values of $X_{ii}$ increase the cost function, while $\emph{\textbf{X}}$s satisfying constraint $\textbf{(3)}$ yield negative contributions, which decrease the cost function.\\
Again, if $N$ is not a power of two, since the transformation $U_d\otimes U_a$ has a trivial action on the states $\{|k\,(\mbox{mod}\,2) \rangle\otimes|k\,(\mbox{mod}\,2) \rangle\}_{k\in\mathcal{B}}$ with $\mathcal{B}=(N,2^n]$, we set $D_{ij}=0$ for all $(i,j)$ such that $i\in (N,2^n]$ or $j\in (N,2^n]$ and we take the intersection of the complementary set of $S$ with the set $\{1\ldots N\}$ in the subtour-elimination term. This makes the states not representing a city irrelevant during the optimization of ${\rm C}$.\\
The cost function in Eq.~\eqref{eq:cost_funct} is linear with respect to $\emph{\textbf{X}}$, and it does not contain terms associated with constraints $\textbf{(1)}$-$\textbf{(2)}$, reported in Eq.~\eqref{eq:class_constraints}, since $\emph{\textbf{X}}$ is a doubly-stochastic matrix by construction.
Moreover, since $\emph{\textbf{X}}$ is a convex combination of classical $\emph{\textbf{x}}$, the total cost function is a convex combination of the costs associated with each classical route satisfying constraints $\textbf{(1)}$-$\textbf{(2)}$:
\begin{equation}
    {\rm C}(\boldsymbol{\alpha}) = \sum_k \lambda_k(\boldsymbol{\alpha}) \, {\rm C}_k
    \quad, \quad \sum_k \lambda_k =1 \,\,\mbox{and}\,\,\lambda_k\ge 0 \,\, \forall k\,,
\end{equation}
where $\{{\rm C}_k\}_k$ are the cost functions associated with the classical routes.
The total cost function ${\rm C}(\boldsymbol{\alpha})$ is limited from below by the cost function value ${\rm C}_{opt}$ associated with the minimal route length satisfying all the constraints in Eq.~\eqref{eq:class_constraints}. Indeed, it holds
\begin{equation}
    {\rm C}(\boldsymbol{\alpha}) \ge \sum_k \lambda_k(\boldsymbol{\alpha}) \, {\rm C}_{opt} = {\rm C}_{opt}\,\sum_k \lambda_k(\boldsymbol{\alpha}) = {\rm C}_{opt} \equiv {\rm C}(\boldsymbol{\alpha}_{opt}) \,,
\end{equation}
where $\boldsymbol{\alpha}_{opt}$ is the optimal configuration of variational parameters such that the quantum route adjacency matrix $\emph{\textbf{X}}(\boldsymbol{\alpha}_{opt})$ is equal to the optimal classical route adjacency matrix $\emph{\textbf{x}}_{opt}$.
As anticipated, the cost function and its minimum do not depend on the specific decomposition of $\emph{\textbf{X}}$, Eq.~\eqref{eq:birkhoff}, since ${\rm C}$ depends only on $\emph{\textbf{X}}$, and so it is uniquely identified by the position on the Birkhoff–von Neumann polytope.
Let us note that the Birkhoff–von Neumann theorem, Eq.~\eqref{eq:birkhoff}, acts analogously to the Rayleigh-Ritz variational principle~\cite{helgaker2014molecular} in the case of the VQE~\cite{Peruzzo_2014,McClean_2016}. Thus, it opens up the possibility of using variational quantum algorithms as solvers for TSPs.

The subtour-elimination term in the cost function contains a number of terms exponentially-growing with respect to $N$ due to the sum over generic subsets $S\subseteq \{1\ldots N\}$ such that $0 < |S| < N$. This means that the evaluation of the overall cost function becomes resource-intensive for the classical machine involved in the VQA for large numbers of cities. A possible strategy to reduce this load consists of starting the cost function in Eq.~\eqref{eq:cost_funct} with $A_{\rm sub}=0$ and adding terms of the sum for subtour elimination if necessary. For example, if the algorithm with $A_{\rm sub}=0$ returns a subtour solution, we identify the subset $S$ for which the solution does not satisfy the constraint $\textbf{(3)}$ and we add only that term to the cost function. In this way, we activate only the needed terms of the subtour-elimination contribution proportional to $A_{\rm sub}$.

At the end of the algorithm, the solution can be read directly from the observable $\emph{\textbf{X}}$ without the need of quantum tomography for the associated optimal trial state $|\psi\rangle$.
In contrast to other approaches~\cite{Lucas_2014,ruan2020quantum,egger2021quantum}, our solution to the TSP is not directly written in the state itself, but in the correlation matrix between the two registers constituting the overall maximally entangled state $|\psi\rangle$ in Eq.~\eqref{eq:final_prep_state}.

\section{Silicon photonics implementation}
\label{sec:experiments}

\begin{figure}[t]
    \centering
    \includegraphics[scale=0.5]{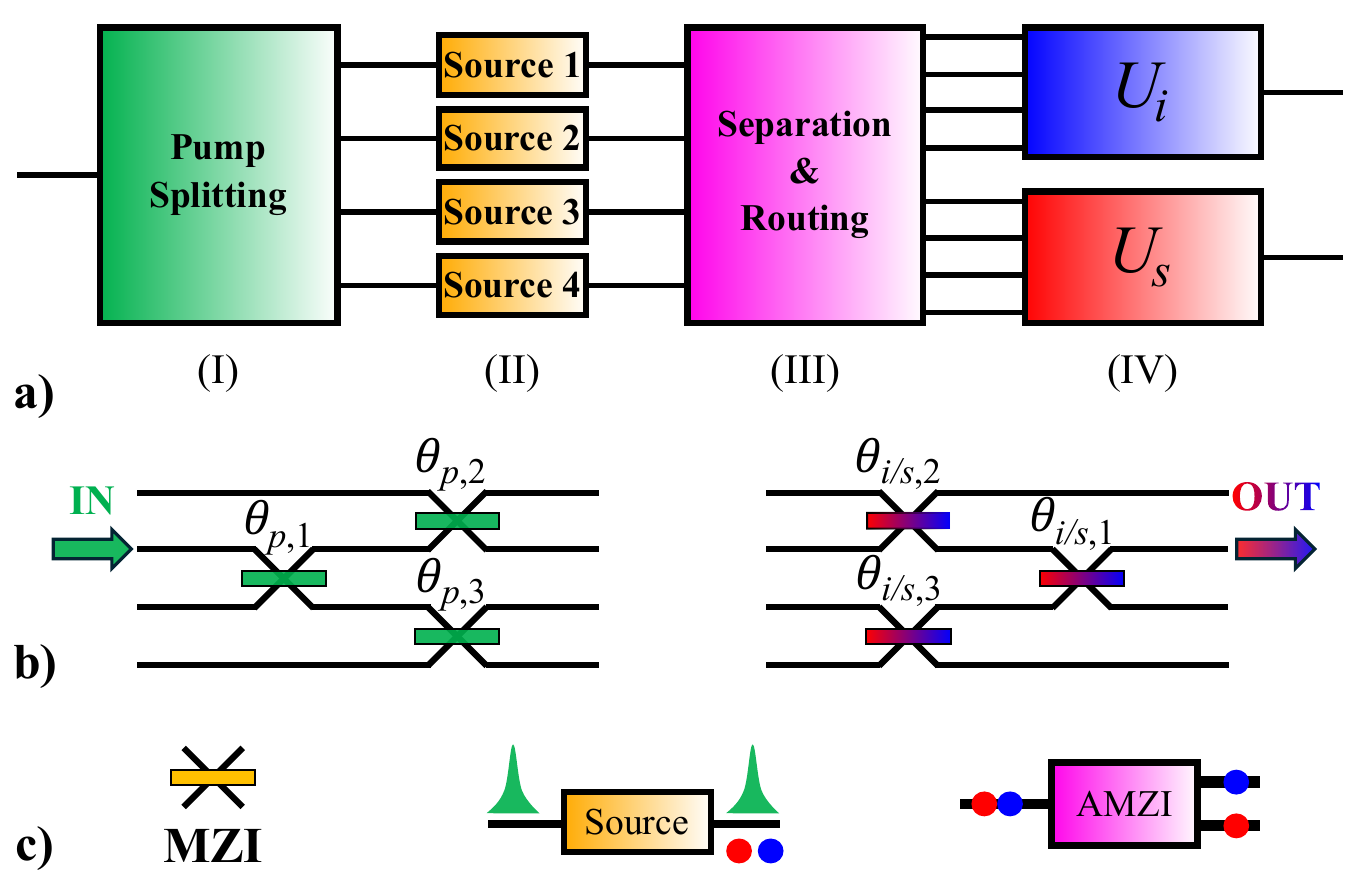}
    \caption{
    \textbf{Block scheme of the photonic circuit able to implement the VQA for four-city TSPs.}
    \textbf{(a)} The Si-PIC is composed of four stages: (I) pump splitting, (II) sources of photon pair, (III) separation and routing, and (IV) independent linear manipulation $U_{i/s}$ of the generated photon pair~\cite{Baldazzi2025}. 
    By equally pumping all four sources, after the stage (III), the state of the two photons can encode two maximally entangled ququarts. In particular, the state can be mapped to the maximally entangled state in Eq.~\eqref{eq:max_ent} with $n=2$. Finally, after stage (IV), the state is precisely the one shown in Eq.~\eqref{eq:final_prep_state}.
    Since we utilized just two outputs and two detectors, one for each component of the bipartite system, the transformation $U_i\otimes U_s$ performs multiple projective measurements equivalent to the action of the transformation $U_d\otimes U_a$ in Figure~\ref{fig:quantum_circuit}.
    \textbf{(b)} Graphical representation of the triangular network of MZIs, contained  in the stage (I) for pump splitting (left) and in the stage (IV) for the linear manipulations $U_{i/s}$ (right).
    \textbf{(c)} From left to right, symbols for MZI, spontaneous-four-wave-mixing-based photon pair source of stage (II) (pump laser in green, signal and idler photons as red and blue dots), and asymmetric MZI (AMZI) of stage (III).}
    \label{fig:phot_circuit}
\end{figure}

The proposed method has been verified by using the Si-PIC described in~\cite{Baldazzi2025}, whose scheme is shown in Figure~\ref{fig:phot_circuit}(a). 
The photonic circuit consists of different stages.
Stage (I) is used to coherently split (see Figure~\ref{fig:phot_circuit}(b) left) a CW laser at 1549.3 nm. In stage (II), four spiral-waveguide-based probabilistic entangled-photon sources are excited with arbitrary amplitudes and generate non-degenerate photon pairs through spontaneous four-wave mixing, where, as usual, the two generated photons are denoted as idler (i) and signal (s), the first with shorter wavelength and the latter with longer wavelength.~\cite{Clemmen_09,Helt_10,Azzini2012,Engin2013} (see Figure~\ref{fig:phot_circuit}(c) center). In the low squeezing regime, the overall state is a spatial superposition of a photon pair with amplitude probabilities related to the setting of the pump splitting. In the experiments, each photon pair source is pumped by an optical power of 0.5 mW and this results in a pair generation rate of about 25 Hz at the detectors. Through asymmetric MZIs and crossing waveguides in stage (III), the twin photons are separated and routed (see Figure~\ref{fig:phot_circuit}(c) right). In particular, this operation converts the energy-time correlation of the photon pair into the spatial correlation for the bipartite system composed of the twin photons, each propagating among four spatial modes, namely paths or waveguides. Then, in stage (IV), linear manipulation of each photon is performed with a triangular MZI scheme (see Figure~\ref{fig:phot_circuit}(b) right). These stages realize the unitary transformations $U_{i}$ and $U_{s}$. Finally, coincidence events of photons are detected using an off-chip single-photon avalanche diode at only one output after each $U_{i}$ and $U_{s}$. The lack of a universal scheme~\cite{reck_experimental_1994,clements_optimal_2016} and the presence of just two detectors among the 8 outputs is compensated by using multiple projective measurements within stage (IV).
More precisely, depending on the desired observable, we choose a set of 16 projectors for the transformation $U_i\otimes U_s$ to reconstruct the expectation value of specific commuting observables.\\
Since the presented bipartite system is composed of two path-encoded ququarts, one for the idler and one for the signal, or equivalently, the quantum register of our system is composed of four qubits, the circuit can be used to prepare trial states for TSPs up to $N=4$. We utilize the idler photon spatial modes for the departure register and the signal photon ones for the arrival register. The correlation matrix $\emph{\textbf{X}}$ results from the spatial correlations of the photon pairs.

To execute our VQA for the TSP, we set the pump splitting to equally excite all four integrated photon pair sources. After stage (III), the photon pair's state can encode the maximally entangled state in Eq.~\eqref{eq:max_ent} with $n=2$~\cite{Baldazzi2025}. Stage (IV) executes the transformation $U_{d}(\boldsymbol{\alpha}_d)\otimes U_a(\boldsymbol{\alpha}_a)$ reported in Figure~\ref{fig:quantum_circuit} through 16 independent projectors, 
\begin{equation}
    \left\{ U_{i}^{(j_i)} \left( \boldsymbol{\theta}_i^{(j_i)}(\boldsymbol{\alpha}_d)\right) \right\}_{j_i\in(1\ldots 4)} \otimes \left\{ U_{s}^{(j_s)}\left( \boldsymbol{\theta}_s^{(j_s)}(\boldsymbol{\alpha}_a)\right) \right\}_{j_s\in(1\ldots 4)} \,,
    \label{eq:vec_pro_comp_var}
\end{equation}
where $\boldsymbol{\alpha} = (\boldsymbol{\alpha}_{d},\boldsymbol{\alpha}_{a})$ are the variational parameters, and $\{\boldsymbol{\theta}_{i/s}^{(j)}\}_{j\in(1\ldots 4)}$ are the phase setting of $U_{i/s}$. The phases $\{\boldsymbol{\theta}_{i/s}^{(j)}\}_{j\in(1\ldots 4)}$ are functions of $\boldsymbol{\alpha}_{d/a}$, in such a way to achieve the same results of $U_{d}\otimes U_{a}$ with 16 multiple projective measurements.
Therefore, we are able to sample the state shown in Eq.~\eqref{eq:final_prep_state} by using 16 runs, each associated with one of the 16 projectors in Eq.~\eqref{eq:vec_pro_comp_var}.
The described setting implies that stages (I) and (III) work with a fixed setting, while the phases of MZIs in stage (IV) are varied during the optimization routine.
The number of variational parameters is six: three for the projectors $\{ U_{i}^{(j_i)} \}_{j_i\in(1\ldots 4)}$ to manipulate the departure register and three for the projectors $\{ U_{s}^{(j_s)} \}_{j_s\in(1\ldots 4)}$ to manipulate the arrival register.
In Appendix~\ref{app:4city}, we show the variational parameters for four-city TSPs with our VQA and how to relate $U_{d/a}(\boldsymbol{\alpha}_{d/a})$ with $\{ U_{i/s}^{(j)}(\boldsymbol{\theta}_{i/s}^{(j)}(\boldsymbol{\alpha}_{d/a})) \}_{j\in(1\ldots 4)}$.

\begin{figure}[t]
    \centering
    \includegraphics[width=\textwidth]{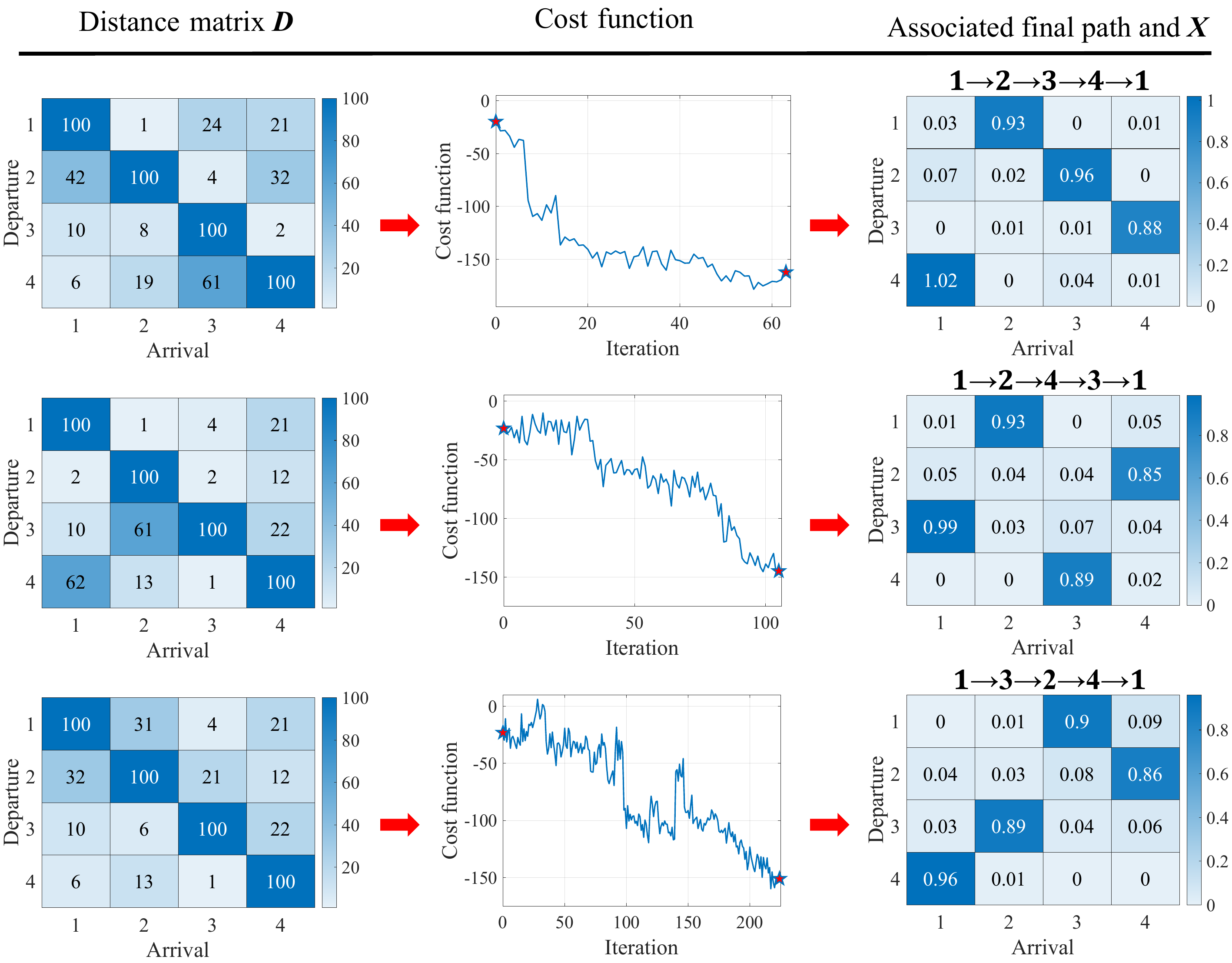}
    \caption{\textbf{Solving four-city TSPs with VQA.}
    In each row, we show the distance matrix $\emph{\textbf{D}}$, the evolution of the minimum gradient-descent search and the final matrix $\emph{\textbf{X}}$ for three examples of TSPs solved with our VQA through our Si-PIC. 
    Above each matrix $\emph{\textbf{X}}$, there is the corresponding route, where starting and ending with city 1 is just a choice.
    We can note that the convergence to the optimal solution is characterized by a different number of iterations: gradient-descend-based optimization algorithms strongly depend on the initial point.
    The entries of the matrix $\emph{\textbf{X}}$ are truncated at the second decimal digit and are affected by 1-2$\%$ relative error.
    }
    \label{fig:test_algo}
\end{figure}

Before starting with the experimental realization of the VQA, the maximally entangled state of dimension four is prepared and its dimensionality is tested using the method explained in~\cite{Sikora_2016,Wang_2018} and implemented in~\cite{Wang_2018,Baldazzi2025}.
We obtain the certified dimension equal to $3.96\pm0.02$, witnessing a nearly ideal degree of entanglement for the produced trial states.\\
Then, the VQA for TSPs is tested on our Si-PIC with different choices of distance matrix $\emph{\textbf{D}}$.
The quantum route adjacency matrix $\emph{\textbf{X}}$, defined in Eq.~\eqref{eq:quantum_BV}, is evaluated taking the coincidence events of the twin photons for each combination of projections:
\begin{equation}
\begin{split}
    & X^{\rm (exp)}_{ij}(\boldsymbol{\alpha}) = 4  
    \frac{{\rm CC}[\boldsymbol{\alpha},i,j] }{{\rm CC}^{\rm tot}[\boldsymbol{\alpha}]} \,,\quad
    \mbox{where}\,\,\,{\rm CC}^{\rm tot}[\boldsymbol{\alpha}] \equiv \sum_{i,j=1}^4
    {\rm CC}[\boldsymbol{\alpha},i,j] \,,
    \end{split}
    \label{eq:XX_CC}
\end{equation}
${\rm CC}$ is the coincidence counts measured with a specific setting of the six variational parameters $\boldsymbol{\alpha}=(\boldsymbol{\alpha}_d,\boldsymbol{\alpha}_a)$ and the $(i,j)$-th projective measurement $U_{i}^{(i)} \left( \boldsymbol{\alpha}_d\right) \otimes U_{s}^{(j)}\left( \boldsymbol{\alpha}_a\right) $, and ${\rm CC}^{\rm tot}$ the total coincidence counts of all projective measurements.
Note that by construction, the entries of $\emph{\textbf{X}}^{\rm (exp)}$ sum up to 4: this property is satisfied by doubly-stochastic matrices.
The estimation of the cost function is executed on a PC by taking the coincidence counts provided by our photonic processor. From these counts, $\emph{\textbf{X}}$ is evaluated with Eq.~\eqref{eq:XX_CC}, and the cost function in Eq.~\eqref{eq:cost_funct} is computed.
In the analysed examples, we choose $D_{ii}=100$ and $A_{\rm sub}=50$: these values ensure that the route with minimum length satisfies the constraints reported $\textbf{(0)}$-$\textbf{(3)}$ in Eq.~\eqref{eq:class_constraints}.

We test our VQA on our Si-PIC with different choices for the distance matrix $\emph{\textbf{D}}$. In all cases, we start by setting the initial ansatz of the six phases $\boldsymbol{\alpha}=(\boldsymbol{\alpha}_d,\boldsymbol{\alpha}_a)$ with random values, and then we execute a gradient-descent-based optimization~\cite{boyd2004convex}. 
As expected, we observed that the convergence of a gradient-based method depends on the initial guess and the learning rate. Moreover, because of the method's features and experimental statistical errors, we can observe slow convergence, convergence to local minima, or even no convergence.
These limits are intrinsically related to the classical optimization problem and the quantum projection noise~\cite{shot_noise}, due to the need for a large number of measurements~\cite{Tilly_2022,Scriva_2024}. Thus, we point out the fact that these limitations are given by the hybrid nature of VQAs and their dependence on a classical optimization, which alone is classically an NP problem~\cite{nesterov2000squared}.

Figure~\ref{fig:test_algo} presents three examples of solved TSPs with our Si-PIC. In all these cases, the optimal solution to the TSP is obtained. In particular, if we evaluate the overlap of the found $\emph{\textbf{X}}^{\rm (exp)}(\boldsymbol{\alpha}_{opt})$ and the solution $\emph{\textbf{x}}_{opt}$ as $4^{-1} \, \mbox{Tr}[\emph{\textbf{X}}^{\rm (exp)}(\boldsymbol{\alpha}_{opt}) \cdot\emph{\textbf{x}}_{opt}^{\rm T}]$, we obtain $\{ 95\%, 91\%, 90\% \}$ for the three examples.
The matrices $\emph{\textbf{X}}^{\rm (exp)}$ are almost doubly-stochastic matrices, and the deviations are due to the non-idealities of the manipulation.
As mentioned previously, we can note that the convergence to the optimal route is reached with a different iteration number, which depends on the initial point and the hyperparameters of the gradient-descent algorithm. The use of gradient-free methods can improve the convergence rate and sensitivity to the initial guess~\cite{Bonet_Monroig_2023}. An example is given by Bayesian optimization~\cite{Iannelli_2022rZ,mockus2005bayesian,bayopt,Garnett_2023}, which, however, is not trivial to generalize to the case of two or more variational parameters.
Because of the multiple projective measurements, the high insertion losses, the low pair generation rate provided by spiral waveguides and the 30\% efficiency of the single-photon detectors, the overall execution times lie approximately between 10 and 30 h. To decrease this time, one can consider the use of four detectors per photon, low-loss waveguides, efficient gratings and brighter sources, such as micro-ring resonators. For example, with superconducting nanowire single-photon detectors~\cite{esmaeil2021superconducting}, ultra-low-loss waveguides~\cite{Bauters_11}, more efficient gratings~\cite{di2026high} and one billion-brightness sources~\cite{chen2024ultralow}, the previous times could be reduced to tens of seconds.

\section{Discussion} 
\label{sec:disc}

In this work, we have introduced a new formulation of VQA for TSPs, and we have successfully demonstrated it on a silicon photonic integrated circuit with integrated photon pair sources. The central ingredients are a maximally entangled state of two registers and its correlation matrix. 
All obtained outcomes validate the proposed solver. This achievement marks a novel application of the currently available photonic technology to relevant and useful combinatorial tasks such as the TSP. 
The disadvantages of our method are given by the number of contributions in the subtour-elimination term and the dependence of the convergence with respect to the utilized gradient-descent-based classical optimization routine. Both aspects are discussed at the end of Section~\ref{sec:var_algorithm} and Section~\ref{sec:experiments}, respectively. Note that this overhead is related to the digital classical part, and thus it is a common feature for any VQA.

\begin{table}[ht]
\centering
\scriptsize
\begin{tabular}{c c c c c}
\toprule
  & QUBO & mQUBO and QDP & SQ & MES \\ 
\midrule
\midrule
\multicolumn{5}{c}{\textbf{Theoretical formulation}}  \\ 
\midrule
References & \cite{Lucas_2014,ruan2020quantum} & \cite{ramezani2024reducing,Glos2022SpaceEfficient,xujun2025quantumspeedupalgorithmtsp} & \cite{goswami2024solving} & This work\\ 
\midrule
Qubit scaling  & $O(N^2)$ & $O\left(N \,\lceil\log_2 N\rceil\right) $ & $1$ & $O(\lceil\log_2 N\rceil)$ \\ 
\midrule
Two-qubit gate scaling  & $O(N^3)$ & $O(N^4 \,\lceil\log_2 N\rceil)$ and $O(N^{5/2})$ & $0$ & $O(N^2)$ \\ 
\midrule
\midrule
\multicolumn{5}{c}{\textbf{Experimental validation}}  \\ 
\midrule
References & \cite{mcgeoch2013experimental,Jain2021_TSP_DWave,stogiannos2022experimental,PadmasolaEtAl2025} & - & - & This work\\ 
\midrule
Hardware & Superconducting & - & - & Photonics \\
\midrule
Number of cities  & $\le12$ & - & - & 4 \\ 
\bottomrule
\end{tabular}
\caption{ \textbf{Comparison among different quantum solvers for the N-city TSP.} 
QUBO stands for quadratic unconstrained binary optimization, mQUBO for modified QUBO, QDP for quantum dynamic programming, SQ for single-qubit and MES for maximally entangled state.
A dash means no experimental realization.
}
\label{tab:comparison}
\end{table}

The new VQA solver 
allows to parallelize the search for the optimal solution among the possible routes and to utilize a reduced number of qubits and gates with respect to the algorithms mentioned in Section~\ref{sec:literature}. 
Indeed, most approaches in the literature treat each entry of the route adjacency matrix $\emph{\textbf{x}}$ as the expectation value of a qubit, and penalty terms are introduced in the cost function in order to satisfy the constraints for an allowed route. In this way, the number of qubits scales as $O\left(N^2\right)$, and the number of two-qubit gates as $O\left(N^3\right)$. It is possible to reduce the qubit number to $O\left(N\, \log_2 N\right)$, but the scaling of the two-qubit gates becomes $O\left(N^4\, \log_2 N\right)$. 
The approaches with $O\left(N^2\right)$-encoding, like in QUBO formulation, are characterized by limited scalability and error sensitivity, and they are still far away from being competitive with classical state-of-the-art TSP solvers.~\cite{smith2025travelling}.
Another possibility consists of using more sophisticated encoding based on ranking and Lehmer codes~\cite{Glos2022SpaceEfficient,BourreauFleuryLacomme2023}, which, however, involves less intuitive forms for the preparation of the trial state and the cost function.
Then, the method with a single qubit~\cite{goswami2024solving} is theoretically interesting, but it does not possess robustness and scalability for practical TSPs~\cite{Baniata_25}.
In our proposal, the number of qubits is $2 \lceil\log_2 N\rceil$. This means that 20 qubits could accommodate up to 1024 cities, a feasible number for state-of-the-art PICs. Problems with cities number in the order of ten thousand could be addressed by means of tens of chips organized in a modular architecture~\cite{aghaee2025scaling}.
We stress the fact that the logarithmic scaling of our quantum register, which can be efficiently stored with a $\mbox{poly}(N)$ cost of bits, does not imply the existence of an asymptotically efficient classical algorithm, because of the overheads given by the classical optimization, present in any VQA.
Regarding the manipulation, if we consider the preparation circuit shown in Figure~\ref{fig:quantum_circuit}, the number of CNOT gates to prepare the maximally entangled state of the two registers, Eq.~\eqref{eq:max_ent}, is $\lceil\log_2 N\rceil$, while the number of CNOT gates to implement $U_d$ and $U_a$ scales as $O\left(N^2\right)$~\cite{barenco1995elementary,rudolph2002rebit,nielsen2003synthesis,mottonen2004quantum,shende2004minimal,mendez2019quantum}.
Therefore, this means that the method described in Section~\ref{sec:var_algorithm} requires the minimum number of qubits and two-qubit gates compared to other methods already reported in the literature (see Table~\ref{tab:comparison}). 
Moreover, we point out that, apart from our work, the only other experimental implementations of quantum algorithms for the TSP have been performed with QUBO formulation on D-Wave superconducting hardware~\cite{mcgeoch2013experimental,Jain2021_TSP_DWave,stogiannos2022experimental,PadmasolaEtAl2025}, while many authors only test their methods on digital quantum simulators~\cite{zhu2022realizablegasbasedquantumalgorithm,qian2023comparative,schnaus2024efficient,He2024_QA_GNN_TSP}. 

We validate the proposed approach on a room-temperature photonic platform for four-city TSPs. Then, we have shown how to produce and utilize a maximally entangled bipartite system made of single photons. 
The state manipulation is not performed by qubit gates as in the circuit of Figure~\ref{fig:quantum_circuit}, but it results in the preparation of the same entangled state needed to sample generic trial states for the VQA solver of the TSP. 
In fact, no photonic CNOT gate is present in our circuit, and the entanglement is created through the energy-time correlation given by the non-linear process of spontaneous four-wave mixing~\cite{energy_time1,energy_time2}. Photonic two-photon gates can only be realized as probabilistic gates through post-selection of qubit-structure-preserving events~\cite{cerf_optical_1998,pittman2001probabilistic,knill_scheme_2001,postsel_CZ,browne2005resource,liu2022universal,li2022quantum,li2022chip,liu2023linear,Kwon_24,baldazzi2024}. 
Our photonic circuit proves that quantum utility does not necessarily require quantum gates and the preparation of exotic quantum states, but it can occur by means of entanglement between two parties~\cite{xue2022variational,science_bn7293}. 
Therefore, entanglement-based specific-purpose quantum hardware could become competitive even for NISQ devices.
Indeed, the scalability and modularity offered by the photonic platform could allow to reach instances with a larger number of cities.
However, as said in~\cite{Baldazzi2025}, our circuit architecture has an exponential scaling of its spatial and temporal resources. Therefore, we need to consider a different approach where we do not rely on a simple scaling of our bipartite system with one photon per part. The VQA explained in Section~\ref{sec:var_algorithm} requires a bipartite maximally entangled trial state. For example, this can be practically achieved with the two parts composed of a multiphoton state~\cite{wang201818} or a continuous-variable state~\cite{jia2025continuous}. Already 14 qubits per part can allow to tackle TSP with more than 16\,000 cities. However, the platform must be able to freely reconfigure the correlation of the two parts. This implies that large instances require not only the preparation of a large entanglement at the beginning, but also programmable multi-qubit transformations, which are needed to explore different configurations of the correlation matrix.

As a last remark, by following the analogy with the variational quantum approach, we note that Eq.~\eqref{eq:classicalBV_super} suggests the following "variational classical algorithm". The permutation $\textbf{P}_{\sigma}$ is substituted with a parametrized linear convex combination of permutation matrices associated with allowed routes. Then, the resulting expression for the modified route adjacency matrix $\emph{\textbf{x}}$ can be inserted in the route length, and an optimization routine is executed. However, the main problem with this approach is the exponential scaling of the parameters. 

\newpage
\section*{Methods}


A TUNICS-BT NetTest Wavelength Tunable CW Laser Diode Source followed by Thorlabs' EDFA100x core-pumped erbium-doped fiber amplifier is used to inject the pump light in the PIC.
The output light from the PIC is filtered by Dense Wavelength Division Multiplexing modules (OpNeti and Precision Microptics), which have 200 GHz bandwidth and 100 GHz FWHM. The DWDM channels 35 (centered at 1549.3 nm), 27 (centered at 1556.7 nm) and 41 (centered at 1544 nm) are used for the pump. signal and idler wavelengths, respectively. Their spectral responses are presented in Supplementary Information Section 3 of~\cite{Baldazzi2025}.

The silicon-on-insulator photonic chip has been fabricated using nanofabrication techniques based on e-beam lithography by SiPhotonic Technologies ApS via a commercial MPW service. The photonic circuit for the VQAs has size of 1.5x5 mm$^2$, and its scheme is described in~\cite{Baldazzi2025}. The silicon waveguide core is $220$ nm-thick and $500$ nm-wide. Grating couplers, MMIs and crossing waveguides have been provided in the foundry's PDKs. The thermal phase shifters are made of titanium, 100 $\mu$m-long and have a mean tuning efficiency of 0.14 rad/mA$^2$. The experiments are performed with the fundamental electric-transverse mode, and the polarization is set by maximizing the transmission through manual fiber polarization controllers because of the different insertion losses of the waveguide modes.
In/out coupling is obtained by a lidless fiber array (Meisu Optics), placed on Thorlabs' 6-Axis NanoMax Stage, whose (X,Y,Z) are connected to Thorlabs' 150 V USB Closed-Loop 3-channel piezo-controller. The propagation losses are about 5 dB/cm, and the coupling losses are about 7 dB/facet.

The detection of the residual pump is performed by Thorlabs' PM100USB and Thorlabs' PDA20CS-EC (InGaAs amplified detector). 
The single photons are detected through two id230 ID-Quantique InGaAs single photon detector modules. Both single-photon detectors work in free-running mode, and they have 30/25$\%$ efficiency, 20/40 $\mu$s deadtime for idler/signal channel, respectively. 
The output counts and coincidences of the detectors are collected and managed by time-tagging electronics (Swabian Instruments) connected to a PC.


Current modules (National Instruments) are linked to a power supply (E3631A 80W Triple Output Power Supply) and provide the currents for the thermal phase shifters of the PIC through a printed board circuit. 
MATLAB's codes manage the experimental setup. In particular, the code runs the self-alignment routine for the piezo-controller, acquires the data coming from the time-tagging electronics, the powermeters and sets the current at the thermal phase shifters.
Moreover, the coincidence counts associated with each phase shifter's setting are collected in one intervals of 30 s until the total number of counts is approximately 750 twofold events, which directly determines the statistical error.
The overall time to calculate one value of the cost function associated with one trial state is given by the time to acquire enough statistics times 16, i.e. the number of projectors.
We used an on-chip power equal to 0.5 mW per excited source. This choice is based on non-linear characterization results.
The overall run time of the analysis summarized in Figure~\ref{fig:test_algo} is around 10-30 hours.
In the gradient-descent optimization algorithm, we use a learning rate equal to 0.001 and a step for the gradient evaluation equal to 0.05 rad.


\newpage

\appendix

\section{Properties of the quantum route adjacency matrix}
\label{app:prop_X}

In this section, we prove the properties of the quantum route adjacency matrix $\emph{\textbf{X}}$.

First of all, let us consider the trial state and its density matrix
\begin{equation}
\begin{split}
    |\psi\rangle &= \frac{1}{2^{n/2}}\sum_{k \in\{0,1\}^n} | \xi_k^{(d)} \rangle_d \otimes | \xi_k^{(a)} \rangle_a
    =\frac{1}{2^{n/2}}\sum_{k \in\{0,1\}^n} U_{d} | k \rangle_d \otimes U_{a} | k \rangle_a \\
    &= \frac{1}{2^{n/2}}\sum_{k,i,j \in\{0,1\}^n} u_{ki}^{(d)} \, u_{kj}^{(a)} \, | ij \rangle \,,\\
    \rho &= |\psi\rangle\langle\psi| =\frac{1}{2^{n}}\sum_{k,i,j \in\{0,1\}^n}\sum_{l,m,n \in\{0,1\}^n} u_{ki}^{(d)} \, u_{kj}^{(a)} \, \bar{u}_{lm}^{(d)} \, \bar{u}_{ln}^{(a)}\, | ij \rangle\langle mn|
    \,,
    \label{eq:trial_state_app}
\end{split}
\end{equation}
where $|ij\rangle = |i\rangle_d\otimes|j\rangle_a$, $u_{jk}^{(d)}$ and $ u_{jk}^{(a)}$ are the $(j,k)$-th components of $U_d$ and $U_a$, respectively, and the bar denotes the complex conjugate. \\
Taking the partial trace with respect to the first register:
\begin{equation}
\begin{split}
    \rho_2 &= \sum_{q \in\{0,1\}^n}\!\!\!  {}_a\langle q | \rho | q \rangle_a \\
    &= \frac{1}{2^{n}}\sum_{q \in\{0,1\}^n}\sum_{k,i,j \in\{0,1\}^n}\sum_{l,m,n \in\{0,1\}^n} u_{ki}^{(d)} \, u_{kj}^{(a)} \, \bar{u}_{lm}^{(d)} \, \bar{u}_{ln}^{(a)}\,\delta_{qj}\delta_{qn}\, | i \rangle\langle m| \\
    &= \frac{1}{2^{n}}\sum_{k,i,j \in\{0,1\}^n}\sum_{l,m \in\{0,1\}^n} \!\!\! u_{ki}^{(d)} \, u_{kj}^{(a)} \, \bar{u}_{lm}^{(d)} \, \bar{u}_{lj}^{(a)}\, | i \rangle\langle m| \\
    &= \frac{1}{2^{n}}\sum_{k,i \in\{0,1\}^n}\sum_{l,m \in\{0,1\}^n} u_{ki}^{(d)}  \, \bar{u}_{lm}^{(d)} \,\delta_{kl} \,| i \rangle\langle m| \\
    &= \frac{1}{2^{n}}\sum_{i \in\{0,1\}^n}\sum_{m \in\{0,1\}^n}  \delta_{im}\, | i \rangle\langle m| 
    = \frac{1}{2^{n}}\sum_{i \in\{0,1\}^n}   | i \rangle\langle i| 
    \,,
\end{split}
\end{equation}
we obtain a maximally mixed state, since the two registers composing $|\psi\rangle$ are maximally entangled.\\
The observable $\emph{\textbf{X}}$ is defined as follows:
\begin{equation}
\begin{split}
    X_{ij}(\boldsymbol{\alpha}) & \equiv 
    2^n \, \mbox{Tr}[ \rho(\boldsymbol{\alpha})\, \hat{P}_{ij} ] \,.
    \label{eq:quantum_BV_app}
\end{split}
\end{equation}
Given $\hat{P}_{ij} = \left(|i \rangle \otimes |j \rangle\right) \,\left(\langle i | \otimes \langle j | \right)$ with $(i,j) \in \{0,1\}^n$, we have
\begin{equation}
\begin{split}
    \mbox{Tr}\left[\rho \, \hat{P}_{ij} \right] 
    &= 
    \frac{1}{2^n} \!\!\! \sum_{k,l \in\{0,1\}^n} \!\!\! \langle i | \xi_k^{(d)} \rangle
    \langle j | \xi_k^{(a)} \rangle
    \langle \xi_l^{(d)} | i \rangle
    \langle \xi_l^{(a)} | j \rangle 
    = \frac{1}{2^n} \!\!\! \sum_{k,l \in\{0,1\}^n} \!\!\!\! u_{ik}^{(d)} \, u_{jk}^{(a)} \,
    \bar{u}_{il}^{(d)} \, \bar{u}_{jl}^{(a)} \,,
\end{split}
\end{equation}
and the following relations hold:
\begin{equation}
\begin{split}
    \sum_{i \in\{0,1\}^n} \!\!\! \mbox{Tr}\left[\rho \, \hat{P}_{ij} \right]  &= 
    \frac{1}{2^n} \!\!\! \sum_{i,k,l \in\{0,1\}^n} \!\!\!\! u_{ik}^{(d)} \, u_{jk}^{(a)} \, \bar{u}_{il}^{(d)} \, \bar{u}_{jl}^{(a)}
    = \frac{1}{2^n} \!\!\! \sum_{k,l \in\{0,1\}^n} \!\!\!\! \delta_{kl}\,u_{jk}^{(a)} \, \bar{u}_{jl}^{(a)} = \frac{1}{2^n} \,,
    \\
    \sum_{j \in\{0,1\}^n} \!\!\! \mbox{Tr}\left[\rho \, \hat{P}_{ij} \right]  &= 
    \frac{1}{2^n} \!\!\! \sum_{j,k,l \in\{0,1\}^n} \!\!\!\! u_{ik}^{(d)} \, u_{jk}^{(a)} \, \bar{u}_{il}^{(d)} \, \bar{u}_{jl}^{(a)}
    = \frac{1}{2^n} \!\!\! \sum_{k,l \in\{0,1\}^n} \!\!\!\! \delta_{kl}\,u_{ik}^{(d)} \, \bar{u}_{il}^{(d)} = \frac{1}{2^n} \,:
\end{split}
\end{equation}
these identities follow from the unitarity of $U_{d/a}$.
Thus, we can note that the correlation matrix satisfies the following properties:
\begin{equation}
\begin{split}
    \sum_{i} X_{ij} &=  \sum_{i,k,l \in\{0,1\}^n} \!\!\! 
    u_{ik}^{(d)} \, u_{jk}^{(a)} \,
    \bar{u}_{il}^{(d)} \, \bar{u}_{jl}^{(a)} 
    =  \sum_{k,l \in\{0,1\}^n} \!\!\!  \delta_{kl}\,u_{jk}^{(a)} \, \bar{u}_{jl}^{(a)}  = 1
    \quad,\, \forall\,j\in[1,N] \,,
    \\
    \sum_{j} X_{ij} &= \sum_{j,k,l \in\{0,1\}^n} \!\!\! u_{ik}^{(d)} \, u_{jk}^{(a)} \,
    \bar{u}_{il}^{(d)} \, \bar{u}_{jl}^{(a)}
    =  \sum_{k,l \in\{0,1\}^n} \!\!\! \delta_{kl}\,u_{ik}^{(d)} \, 
    \bar{u}_{il}^{(d)} = 1
    \quad,\, \forall\,i\in[1,N] \,.
\end{split}
\end{equation}
This means that the quantum route adjacency matrix $\emph{\textbf{X}}$ is a doubly-stochastic matrix, like the generic permutation matrix $\emph{\textbf{x}}$ representing a route of TSP.
Therefore, $\emph{\textbf{X}}$ satisfies constraints $\textbf{(1)}$-$\textbf{(2)}$, reported in Section~\ref{sec:class_form}, by construction.

If the state has the following form:
\begin{equation}
    |\psi_{\sigma}\rangle = \frac{1}{2^{n/2}}\sum_{k \in\{0,1\}^n} | k \rangle \otimes | \sigma(k) \rangle \,,
    \label{eq:route_quantum_state_app}
\end{equation}
where $\sigma$ is a permutation, then the associated correlation matrix is simply
\begin{equation}
    X_{ij} = \delta_{\sigma(i),j} \,,
\end{equation}
thus, the permutation matrix corresponding to $\sigma$.
On the other hand, if the correlation matrix is a permutation matrix and we consider pure states of the form \eqref{eq:trial_state_app}, then the state has the form reported in Eq.~\eqref{eq:route_quantum_state_app}, modulo generic relative phases among the computational basis terms composing it. 
More precisely, the proof follows from the following steps:
\begin{equation}
\begin{split}
    & \delta_{\sigma(i),j} = \sum_{k, l \in\{0,1\}^n} \!\!\!\! u_{ki}^{(d)} \, u_{kj}^{(a)} \, \bar{u}_{li}^{(d)} \, \bar{u}_{lj}^{(a)} \,, \\
    & \left[U_{d}\otimes U_{a}\right] | \psi_0 \rangle = \left[\mathbf{1}\otimes \tilde{U}_{a} \right]| \psi_0 \rangle \implies u_{ij}^{(d)} = \delta_{ij} \quad \mbox{modulo phases w/o loss of generality}\,, \\
    & \delta_{\sigma(i),j} = \sum_{k, l \in\{0,1\}^n} \!\!\!\! \delta_{ki} \, u_{kj}^{(a)} \, \delta_{li} \, \bar{u}_{lj}^{(a)} = \left| u_{ij}^{(a)} \right|^2 \implies u_{ij}^{(a)} = \delta_{\sigma(i),j} \quad \mbox{modulo phases} \,,
\end{split}
\end{equation}
where the second line is a consequence of the property of maximally entangled states~\cite{khatri2019quantum,xue2022variational}.

Therefore, modulo relative phases, a correlation matrix equal to a permutation matrix corresponds to a state given in Eq.~\eqref{eq:route_quantum_state_app}.


\section{VQA setting for four-city TSPs}
\label{app:4city}

In this section, we describe the general features of the classical four-city TSP, and then we show how to prepare the trial state for our VQA in the general case and in the case of our photonic circuit. In particular, we discuss the chosen form of $U_d\otimes U_a$, which removes some redundancy of the problem and reduces the number of variational parameters to six, and the associated transformations $U_i\otimes U_s$, which achieve the same result as $U_d\otimes U_a$ through multiple projective measurements.

\subsection{Four-city routes}

The four-city TSP has six possible routes that satisfy all the constraints reported in Section~\ref{sec:class_form}. The matrices $\emph{\textbf{x}}$ representing these routes are:
\begin{equation}
    \begin{split}
        &   \begin{pmatrix}
            0 & 1 & 0 & 0 \\
            0 & 0 & 1 & 0 \\
            0 & 0 & 0 & 1 \\
            1 & 0 & 0 & 0 
            \end{pmatrix} \iff \textbf{1}\to\textbf{2}\to\textbf{3}\to\textbf{4}\to\textbf{1}
        \quad,\quad
            \begin{pmatrix}
            0 & 0 & 1 & 0 \\
            0 & 0 & 0 & 1 \\
            0 & 1 & 0 & 0 \\
            1 & 0 & 0 & 0 
            \end{pmatrix} \iff \textbf{1}\to\textbf{3}\to\textbf{2}\to\textbf{4}\to\textbf{1} \quad,
            \\
        &   \begin{pmatrix}
            0 & 0 & 0 & 1 \\
            1 & 0 & 0 & 0 \\
            0 & 1 & 0 & 0 \\
            0 & 0 & 1 & 0 
            \end{pmatrix} \iff \textbf{1}\to\textbf{4}\to\textbf{3}\to\textbf{2}\to\textbf{1}
        \quad,\quad
            \begin{pmatrix}
            0 & 1 & 0 & 0 \\
            0 & 0 & 0 & 1 \\
            1 & 0 & 0 & 0 \\
            0 & 0 & 1 & 0 
            \end{pmatrix} \iff \textbf{1}\to\textbf{2}\to\textbf{4}\to\textbf{3}\to\textbf{1} \quad,
            \\
        &   \begin{pmatrix}
            0 & 0 & 0 & 1 \\
            0 & 0 & 1 & 0 \\
            1 & 0 & 0 & 0 \\
            0 & 1 & 0 & 0 
            \end{pmatrix} \iff \textbf{1}\to\textbf{4}\to\textbf{2}\to\textbf{3}\to\textbf{1}
        \quad,\quad
            \begin{pmatrix}
            0 & 0 & 1 & 0 \\
            1 & 0 & 0 & 0 \\
            0 & 0 & 0 & 1 \\
            0 & 1 & 0 & 0 
            \end{pmatrix} \iff \textbf{1}\to\textbf{3}\to\textbf{4}\to\textbf{2}\to\textbf{1} \quad,
    \end{split}
\label{eq:4city_allowed}
\end{equation}
where the associated path is shown on the right-hand side of each matrix.
There are also three routes with subtours, or equivalently routes which don't satisfy only the constraint $\textbf{(3)}$. The associated matrices $\emph{\textbf{x}}$ are:
\begin{equation}
    \begin{split}
        &   \begin{pmatrix}
            0 & 1 & 0 & 0 \\
            1 & 0 & 0 & 0 \\
            0 & 0 & 0 & 1 \\
            0 & 0 & 1 & 0 
            \end{pmatrix} \iff \textbf{1}\leftrightarrow\textbf{2}\cup \textbf{3}\leftrightarrow\textbf{4}
        \quad,\quad
            \begin{pmatrix}
            0 & 0 & 1 & 0 \\
            0 & 0 & 0 & 1 \\
            1 & 0 & 0 & 0 \\
            0 & 1 & 0 & 0 
            \end{pmatrix} \iff \textbf{1}\leftrightarrow\textbf{3}\cup \textbf{2}\leftrightarrow\textbf{4} \quad,
            \\
        &   \begin{pmatrix}
            0 & 0 & 0 & 1 \\
            0 & 0 & 1 & 0 \\
            0 & 1 & 0 & 0 \\
            1 & 0 & 0 & 0 
            \end{pmatrix} \iff \textbf{1}\leftrightarrow\textbf{4}\cup \textbf{2}\leftrightarrow\textbf{3}
            \quad,
    \end{split}
    \label{eq:4city_not_allowed}
\end{equation}
where again the associated path is shown on the right-hand side of each matrix.

\subsection{Trial state preparation}

First of all, let us consider the states $|\psi\rangle$, reported in Eq.~\eqref{eq:trial_state_app}, whose correlation matrices $\emph{\textbf{X}}$ give the six routes for the four-city case, shown in Eq.~\eqref{eq:4city_allowed}.\\
These states read as follows
\begin{equation}
\begin{split}
    |\psi^{(4)}_{1234}\rangle &= \frac{1}{2} \left[ 
    | 00 \rangle_d \otimes | 01 \rangle_a +
    | 01 \rangle_d \otimes | 10 \rangle_a +
    | 10 \rangle_d \otimes | 11 \rangle_a +
    | 11 \rangle_d \otimes | 00 \rangle_a
    \right] \,,
    \\
    |\psi^{(4)}_{1243}\rangle &= \frac{1}{2} \left[ 
    | 00 \rangle_d \otimes | 01 \rangle_a +
    | 01 \rangle_d \otimes | 11 \rangle_a +
    | 10 \rangle_d \otimes | 00 \rangle_a +
    | 11 \rangle_d \otimes | 10 \rangle_a
    \right] \,,
    \\
    |\psi^{(4)}_{1324}\rangle &= \frac{1}{2} \left[ 
    | 00 \rangle_d \otimes | 10 \rangle_a +
    | 01 \rangle_d \otimes | 11 \rangle_a +
    | 10 \rangle_d \otimes | 01 \rangle_a +
    | 11 \rangle_d \otimes | 00 \rangle_a
    \right] \,,
    \\
    |\psi^{(4)}_{1342}\rangle &= \frac{1}{2} \left[ 
    | 00 \rangle_d \otimes | 10 \rangle_a +
    | 01 \rangle_d \otimes | 00 \rangle_a +
    | 10 \rangle_d \otimes | 11 \rangle_a +
    | 11 \rangle_d \otimes | 01 \rangle_a
    \right] \,,
    \\
    |\psi^{(4)}_{1423}\rangle &= \frac{1}{2} \left[ 
    | 00 \rangle_d \otimes | 11 \rangle_a +
    | 01 \rangle_d \otimes | 10 \rangle_a +
    | 10 \rangle_d \otimes | 00 \rangle_a +
    | 11 \rangle_d \otimes | 01 \rangle_a
    \right] \,,
    \\
    |\psi^{(4)}_{1432}\rangle &= \frac{1}{2} \left[ 
    | 00 \rangle_d \otimes | 11 \rangle_a +
    | 01 \rangle_d \otimes | 00 \rangle_a +
    | 10 \rangle_d \otimes | 01 \rangle_a +
    | 11 \rangle_d \otimes | 10 \rangle_a
    \right] \,,
\end{split}
\end{equation}
where the superscript denotes the number of cities and the subscript the route.\\
Analogously, the states, whose correlation matrices $\emph{\textbf{X}}$ give the three routes with subtours in Eq.~\eqref{eq:4city_not_allowed} are:
\begin{equation}
\begin{split}
    |\psi^{(4)}_{12-34}\rangle &= \frac{1}{2} \left[ 
    | 00 \rangle_d \otimes | 01 \rangle_a +
    | 01 \rangle_d \otimes | 00 \rangle_a +
    | 10 \rangle_d \otimes | 11 \rangle_a +
    | 11 \rangle_d \otimes | 10 \rangle_a
    \right] \,,
    \\
    |\psi^{(4)}_{13-24}\rangle &= \frac{1}{2} \left[ 
    | 00 \rangle_d \otimes | 10 \rangle_a +
    | 10 \rangle_d \otimes | 00 \rangle_a +
    | 01 \rangle_d \otimes | 11 \rangle_a +
    | 11 \rangle_d \otimes | 01 \rangle_a
    \right] \,,
    \\
    |\psi^{(4)}_{14-23}\rangle &= \frac{1}{2} \left[ 
    | 00 \rangle_d \otimes | 11 \rangle_a +
    | 11 \rangle_d \otimes | 00 \rangle_a +
    | 01 \rangle_d \otimes | 10 \rangle_a +
    | 10 \rangle_d \otimes | 01 \rangle_a
    \right] \,.
\end{split}
\end{equation}

It is important to note that phases among the different terms of the previous states do not alter the correlation matrix between the two registers. This means that the unitary transformation $U_d\otimes U_a$ can be chosen to be a real symmetric map, i.e. $\left(U_d\otimes U_a\right)^{\rm T}=U_d\otimes U_a$ with T denoting the transposition.

Now, we analyse the transformation $U_d\otimes U_a$ in state $|\psi\rangle$, Eq.~\eqref{eq:trial_state_app}, in order to parametrize the trial states and identify the variational parameters.\\
We decide to decompose the transformations $U_{d/a}$ in terms of transformations $u \in {\rm SU}(2)$, whose matrix representation reads 
\begin{equation}
u[\Theta, \boldsymbol{\Phi}] \equiv \left(
\begin{array}{cc}
 e^{i \Phi_1 } \sin \Theta  & \quad e^{i \Phi_2 }\cos \Theta \\
 e^{-i \Phi_2 } \cos \Theta  & \quad-e^{-i \Phi_1 }\sin \Theta \\
\end{array}
\right) \quad,\quad
\mbox{where}\,\,\boldsymbol{\Phi} =(\Phi_1,\Phi_2) \,.
\label{eq:su2_matrix}
\end{equation}
Following the decomposition given in~\cite{clements_optimal_2016} and the symmetry requirement, the transformation of one register can be written as follows
\begin{equation}
\begin{split}
    &U[\theta_1, \theta_2, \theta_3, \theta_4, \theta_5, \theta_6] \\
  & = u^{(1,2)}[\theta_1, \boldsymbol{0}] \cdot u^{(3,4)}[\theta_2, \boldsymbol{0}] \cdot u^{(2,3)}[\theta_3, \boldsymbol{0}] \cdot 
   u^{(1,2)}[\theta_4, \boldsymbol{0}] \cdot u^{(3,4)}[\theta_5, \boldsymbol{0}] \cdot u^{(2,3)}[\theta_6, \boldsymbol{0}] \,,
   \label{eq:decom_uni}
\end{split}
\end{equation}
where the superscript denotes the pairs of states on which $u$ is acting. Note that the states are written in base ten just for convenience, but one can recover the qubit states by converting the two numbers in the superscript in base two. 
This decomposition of a generic unitary matrix in ${\rm SU}(2)$ sub-matrices is concretely equivalent to creating a generic $m\times m$ photonic device assembling $2\times2$ photonic devices. Each individual transformation $u^{(k,k+1)}$ is leaving unaffected all inputs different from $k$ and $(k+1)$ ones, which are evolving accordingly to Eq.~\eqref{eq:su2_matrix}.

\begin{figure}[t]
    \centering
    \includegraphics[width=\textwidth]{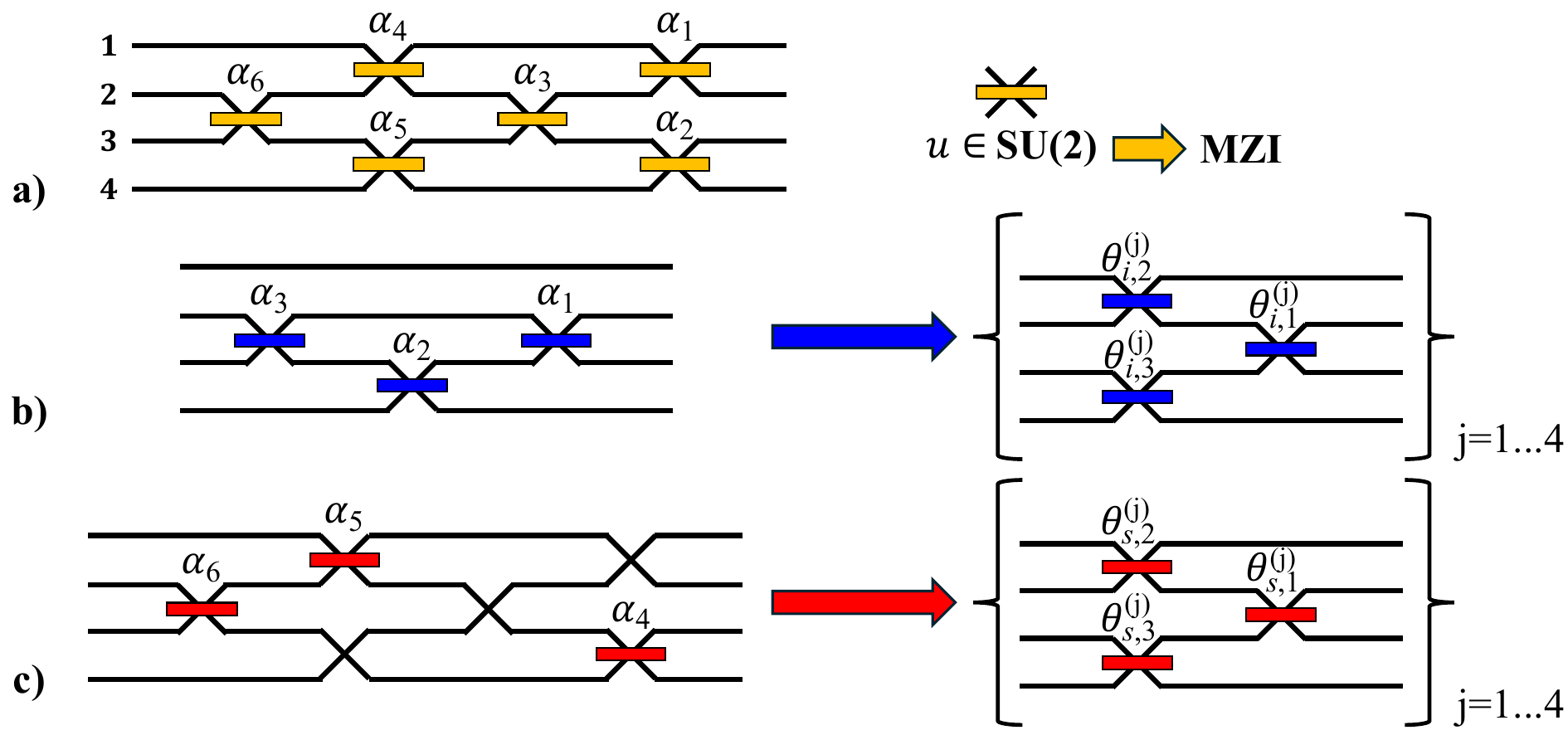}
    \caption{\textbf{Four-dimensional unitary transformation and its equivalent set of projective measurements.}
    \textbf{(a)} Decomposition of a unitary transformation of dimension four~\cite{clements_optimal_2016} in terms of SU$(2)$ matrices given in Eq.~\eqref{eq:su2_matrix}. On the left of the scheme we can find the spatial modes labels, while on the right the symbol for an MZI, whose action is described by an SU$(2)$ matrix.
    \textbf{(b)} On the left, utilized decomposition of a unitary transformation $U_d$ satisfying the condition in Eq.~\eqref{eq:cond_UU_app}(left). On the right, the triangular decomposition whose action is equivalent through multiple projective measurements, i.e. $\left\{U_i^{(j)}\right\}_{j\in (1\ldots 4)}$.
    \textbf{(c)} On the left, utilized decomposition of a unitary transformation $U_a$ satisfying the condition in Eq.~\eqref{eq:cond_UU_app}(right). On the right, the triangular decomposition whose action is equivalent through multiple projective measurements, i.e. $\left\{U_s^{(j)}\right\}_{j\in (1\ldots 4)}$.
    }
    \label{fig:scheme_photo}
\end{figure}

Figure~\ref{fig:scheme_photo}(a) shows the decomposition in Eq.~\eqref{eq:decom_uni}. In particular, in the context of photonic circuits, the $u\in {\rm SU}(2)$ is implemented with an MZI, as shown on the right, and the MZI network on the left is a 4x4 universal scheme~\cite{reck_experimental_1994,clements_optimal_2016}, since it can execute the generic unitary transformation of four spatial modes.

As we point out in the manuscript, there is redundancy in the state manipulation because of the cyclicity of the TSP.
In order to remove this redundancy and reduce the number of variational parameters entering the map $U_d\otimes U_a$, we choose city 1 as the first departure city and, thus, also the last arrival city.
This choice can be realized by the following requirements: 
\begin{equation}
    U_d \, | 1\rangle = | 1\rangle
    \quad \mbox{and} \quad
    U_a \, | 4\rangle = | 1\rangle \,,
    \label{eq:cond_UU_app}
\end{equation}
where the states are written in base ten just for convenience, but one can recover the qubit states by converting the numbers in base two. Here, we are assuming that the first term of the departure register in the trial state $|\psi\rangle$ is $|00\rangle_d$ and the last term of the arrival register in the trial state $|\psi\rangle$ is $|00\rangle_a$. In this way, the redundancy due to the cyclicity is removed by fixing certain contributions and the summation order in the trial state.
Using the decomposition in Eq.~\eqref{eq:decom_uni}, we can fix $\left\{\theta_1 = \frac{\pi }{2},\theta_2 = \frac{\pi }{2},\theta_4 = \frac{\pi }{2}\right\}$ for $U_d$ and $\left\{\theta_1 = 0,\theta_3 = 0, \theta_5 = 0\right\}$ for $U_a$ to achieve the conditions in the previous equation. This choice is not unique.
Within these assumptions and relabeling the free parameters, the matrix representation for $U_d$ reads as follows
\begin{equation}
\begin{split}
    & U_d[\alpha_1, \alpha_2, \alpha_3] \equiv
U[\pi/2, \pi/2, \alpha_1, \pi/2, \alpha_2, \alpha_3]
\\
& =
\left(
\begin{array}{cccc}
 1 & 0 & 0 & 0 \\
 0 & \sin\alpha_1 \sin\alpha_3-\cos\alpha_1 \sin\alpha_2 \cos\alpha_3 & \sin\alpha_1 \cos\alpha_3+\cos\alpha_1 \sin\alpha_2 \sin\alpha_3 & -\cos\alpha_1 \cos\alpha_2 \\
 0 & -\cos\alpha_1 \sin\alpha_3-\sin\alpha_1 \sin\alpha_2 \cos\alpha_3 & -\cos\alpha_1 \cos\alpha_3+\sin\alpha_1 \sin\alpha_2 \sin\alpha_3 & -\sin\alpha_1 \cos\alpha_2 \\
 0 & -\cos\alpha_2 \cos\alpha_3 & \cos\alpha_2 \sin\alpha_3 & \sin\alpha_2 \\
\end{array}
\right)  \,,
\label{eq:Ud_4case}
\end{split}
\end{equation}
and for $U_a$ as follows
\begin{equation}
\begin{split}
& U_a[\alpha_4, \alpha_5, \alpha_6]=
U[0, \alpha_4, 0, \alpha_5, 0, \alpha_6] \\
& =
\left(
\begin{array}{cccc}
 0 & 0 & 0 & 1 \\
 \sin\alpha_5 & \cos\alpha_5 \sin\alpha_6 & \cos\alpha_5 \cos\alpha_6 & 0 \\
 \sin\alpha_4 \cos\alpha_5 & \cos\alpha_4 \cos\alpha_6-\sin\alpha_4 \sin\alpha_5 \sin\alpha_6
   & -\cos\alpha_4 \sin\alpha_6-\sin\alpha_4 \sin\alpha_5 \cos\alpha_6 & 0 \\
 \cos\alpha_4 \cos\alpha_5 & -\sin\alpha_4 \cos\alpha_6-\cos\alpha_4 \sin\alpha_5 \sin\alpha_6 & \sin\alpha_4 \sin\alpha_6-\cos\alpha_4 \sin\alpha_5 \cos\alpha_6 & 0 \\
\end{array}
\right) \,.
\label{eq:Ua_4case}
\end{split}
\end{equation}
Figure~\ref{fig:scheme_photo}(b)-(c) on the left of the arrow show the previous decompositions of $U_d$ and $U_a$ respectively, satisfying the conditions in Eq.~\eqref{eq:cond_UU_app} and shown in Eq.~\eqref{eq:Ud_4case} and Eq.~\eqref{eq:Ua_4case}. In the context of photonic circuits, these schemes are associated with two 4x4 MZI networks: the first, Figure~\ref{fig:scheme_photo}(b) on the left, maps the first input spatial mode to the first output spatial modes and freely manipulates the other modes, and the second, Figure~\ref{fig:scheme_photo}(c) on the left, maps the fourth input spatial mode to the first output spatial modes and freely manipulates the other modes.

\begin{table}[ht]
\centering
\scriptsize
\begin{tabular}{c ccc || c ccc }
\toprule
Sequence of departures & $\alpha_1$ & $\alpha_2$ & $\alpha_3$ & Sequence of arrivals & $\alpha_4$ & $\alpha_5$ & $\alpha_6$ \\  
\midrule
\midrule
$\textbf{1}\to\textbf{2}\to\textbf{3}\to\textbf{4}$ & $\frac{\pi}{2}$ & $\frac{\pi}{2}$ & $\frac{\pi}{2}$ 
& $\textbf{2}\to\textbf{3}\to\textbf{4}\to\textbf{1}$ & $\frac{\pi}{2}$ & $\frac{\pi}{2}$ & $\frac{\pi}{2}$ \\ 
\midrule
$\textbf{1}\to\textbf{2}\to\textbf{4}\to\textbf{3}$ & $\frac{\pi}{2}$ & $0$ & $\frac{\pi}{2}$ 
& $\textbf{2}\to\textbf{4}\to\textbf{3}\to\textbf{1}$ & $\frac{\pi}{2}$ & $\frac{\pi}{2}$ & $0$ \\  
\midrule
$\textbf{1}\to\textbf{3}\to\textbf{2}\to\textbf{4}$ & $\frac{\pi}{2}$ & $\frac{\pi}{2}$ & $0$ 
& $\textbf{3}\to\textbf{2}\to\textbf{4}\to\textbf{1}$ & $\frac{\pi}{2}$ & $0$ & $\frac{\pi}{2}$ \\  
\midrule
$\textbf{1}\to\textbf{3}\to\textbf{4}\to\textbf{2}$ & $0$ & $0$ & $\frac{\pi}{2}$ 
& $\textbf{3}\to\textbf{4}\to\textbf{2}\to\textbf{1}$ & $\frac{\pi}{2}$ & $0$ & $0$ \\  
\midrule
$\textbf{1}\to\textbf{4}\to\textbf{2}\to\textbf{3}$ & $\frac{\pi}{2}$ & $0$ & $0$ 
& $\textbf{4}\to\textbf{2}\to\textbf{3}\to\textbf{1}$ & $0$ & $0$ & $\frac{\pi}{2}$ \\ 
\midrule
$\textbf{1}\to\textbf{4}\to\textbf{3}\to\textbf{2}$ & $0$ & $0$ & $0$ 
& $\textbf{4}\to\textbf{3}\to\textbf{2}\to\textbf{1}$ & $0$ & $0$ & $0$\\ 
\bottomrule
\end{tabular}
\caption{ \textbf{Phase setting associated with the different sequences of departures and arrivals for allowed routes.}
On the left, the sequence of departures and the choice of phases for the decomposition of $U_d$ shown in Eq.~\eqref{eq:Ud_4case} and Figure~\ref{fig:scheme_photo}(b) on the left.
On the right, the sequence of arrivals and the choice of phases for the decomposition of $U_a$ shown in Eq.~\eqref{eq:Ua_4case} and Figure~\ref{fig:scheme_photo}(c) on the left.
As we pointed out in the main text, these choices of phases are not unique.
}
\label{tab:Ud_Ua_setting}
\end{table}
Table~\ref{tab:Ud_Ua_setting} shows the phases to obtain the sequences of departures and arrivals using the matrix decomposition in Eq.~\ref{eq:Ud_4case} and Eq.~\ref{eq:Ua_4case}. Note that these settings are not unique: different phases can correspond to the same city sequence.

Then, the trial state is the maximally entangled state of the two registers transformed with $U_d\otimes U_a$:
\begin{equation}
\begin{split}
    \psi^{(4)}[ \alpha_1, \alpha_2, \alpha_3, \alpha_4, \alpha_5, \alpha_6] &= 
    U_d[ \alpha_1, \alpha_2, \alpha_3]\otimes U_a[ \alpha_4, \alpha_5, \alpha_6] \cdot \psi_0^{(4)} \,, \\
    \mbox{where} \quad \psi_0^{(4)}&= \frac{1}{2} ( 1, 0, 0, 0, 
    0, 1, 0, 0, 0, 0, 1, 0, 0, 0, 0, 1 ) \,,
\end{split}
\end{equation}
and we have written $|\psi_0\rangle = \frac{1}{2}\sum_{k \in\{0,1\}^2} | k \rangle_d \otimes | k \rangle_a $ in the matrix representation.
The parameters $\left( \alpha_1, \alpha_2, \alpha_3, \alpha_4, \alpha_5, \alpha_6 \right)$ are the variational parameter of our VQA for TSPs.
%
%
Finally, by using the decomposition of $U_d$ shown in Eq.~\eqref{eq:Ud_4case} and the decomposition of $U_a$ shown in Eq.~\eqref{eq:Ua_4case}, the correlation matrix $\emph{\textbf{X}}^{(4)}$ reads
\begin{equation}
\begin{split}
X_{11}^{(4)} &= 0 \,,\\
X_{12}^{(4)} &= \sin ^2(\alpha_5) \,,\\
X_{13}^{(4)} &= \sin ^2(\alpha_4) \cos ^2(\alpha_5) \,,\\
X_{14}^{(4)} &= \cos ^2(\alpha_4) \cos ^2(\alpha_5) \,,\\
X_{21}^{(4)} &= \cos ^2(\alpha_1) \cos ^2(\alpha_2) \,,\\
X_{22}^{(4)} &= \cos ^2(\alpha_5) \left[\sin (\alpha_1) \cos (\alpha_6-\alpha_3)-\cos (\alpha_1) \sin (\alpha_2) \sin (\alpha_6-\alpha_3)\right]^2 \,,\\
X_{23}^{(4)} &= \left\{\left[\cos (\alpha_4) \cos (\alpha_6)-\sin (\alpha_4) \sin (\alpha_5) \sin (\alpha_6)\right] \left[\sin (\alpha_1) \sin (\alpha_3)-\cos (\alpha_1) \sin (\alpha_2) \cos (\alpha_3)\right] \right. \,,\\
&\qquad \left.-\left[\cos
   (\alpha_4) \sin (\alpha_6)+\sin (\alpha_4) \sin (\alpha_5) \cos (\alpha_6)\right] \left[\sin (\alpha_1) \cos (\alpha_3)+\cos (\alpha_1) \sin (\alpha_2) \sin (\alpha_3)\right]\right\}^2 \,,\\
X_{24}^{(4)} &= \left\{\cos (\alpha_6-\alpha_3) \left[\sin (\alpha_4) \cos (\alpha_1) \sin (\alpha_2)-\cos (\alpha_4) \sin (\alpha_5) \sin
   (\alpha_1)\right] \right.\,,\\
&\qquad \left.+\sin (\alpha_6-\alpha_3) \left[\cos (\alpha_4) \sin (\alpha_5) \cos (\alpha_1) \sin (\alpha_2)+\sin
   (\alpha_4) \sin (\alpha_1)\right]\right\}^2 \,,\\
X_{31}^{(4)} &= \sin ^2(\alpha_1) \cos ^2(\alpha_2) \,,\\
X_{32}^{(4)} &= \cos ^2(\alpha_5) \left[\sin (\alpha_1) \sin (\alpha_2) \sin (\alpha_6-\alpha_3)+\cos (\alpha_1) \cos (\alpha_6-\alpha_3)\right]^2 \,,\\
X_{33}^{(4)} &= \left\{\left[\sin (\alpha_4) \sin (\alpha_5) \cos (\alpha_6)+\cos (\alpha_4) \sin (\alpha_6)\right] \left[\cos (\alpha_1) \cos (\alpha_3)-\sin (\alpha_1) \sin (\alpha_2) \sin (\alpha_3)\right] \right.\,,\\
&\qquad \left.-\left[\cos (\alpha_4) \cos (\alpha_6)-\sin (\alpha_4) \sin
   (\alpha_5) \sin (\alpha_6)\right] \left[\sin (\alpha_1) \sin (\alpha_2) \cos (\alpha_3)+\cos (\alpha_1) \sin (\alpha_3)\right]\right\}^2 \,,\\
X_{34}^{(4)} &= \left\{\cos (\alpha_6-\alpha_3) \left[\cos (\alpha_4) \sin (\alpha_5) \cos (\alpha_1)+\sin (\alpha_4) \sin (\alpha_1) \sin
   (\alpha_2)\right] \right.\,,\\
&\qquad \left. +\sin (\alpha_6-\alpha_3) \left[\cos (\alpha_4) \sin (\alpha_5) \sin (\alpha_1) \sin (\alpha_2)-\sin
   (\alpha_4) \cos (\alpha_1)\right]\right\}^2 \,,\\
X_{41}^{(4)} &= \sin ^2(\alpha_2) \,,\\
X_{42}^{(4)} &= \cos ^2(\alpha_5) \cos ^2(\alpha_2) \sin ^2(\alpha_6-\alpha_3) \,,\\
X_{43}^{(4)} &= \cos ^2(\alpha_2) \left[\sin (\alpha_4) \sin (\alpha_5) \sin (\alpha_6-\alpha_3)-\cos (\alpha_4) \cos (\alpha_6-\alpha_3)\right]^2 \,,\\
X_{44}^{(4)} &=  \cos ^2(\alpha_2) \left[\cos (\alpha_4) \sin (\alpha_5) \sin (\alpha_6-\alpha_3)+\sin (\alpha_4) \cos (\alpha_6-\alpha_3)\right]^2
\,.
\end{split}
\label{eq:X_4case_app}
\end{equation}

In our photonic circuit, the manipulation is achieved with a triangular scheme~\cite{Baldazzi2025}. Thus, we don't have universal schemes like the one reported in Figure~\ref{fig:scheme_photo}(a).
However, using multiple projective measurements, it is possible to achieve the same action of a universal scheme.
The triangular scheme can be decomposed in terms of matrices given in Eq.~\eqref{eq:su2_matrix} as follows
\begin{equation}
    U_{triangular}[ \theta_1,\theta_2,\theta_3] = 
  u^{(2,3)}[\theta_1, \boldsymbol{0}] \cdot u^{(1,2)}[\theta_2, \boldsymbol{0}] \cdot u^{(3,4)}[\theta_3, \boldsymbol{0}] \,,
   \label{eq:decom_triangular}
\end{equation}
and it is represented in Figures~\ref{fig:scheme_photo}(b)-(c) on the right. Note that, like the case reported in Eq.~\eqref{eq:decom_uni}, we set to zero all the $\boldsymbol{\phi}$ phases because of the requirement to have real symmetric transformations.

Figures~\ref{fig:scheme_photo}(b)-(c) present the idea of mapping the action of $U_d\otimes U_a$, implemented by universal schemes, into the action of multiple transformations $U_i\otimes U_s$, implemented with a triangular scheme. More precisely, we need to find the following relation
\begin{equation}
\begin{split}
U_d \left( \boldsymbol{\alpha}_d\right)  &\otimes  U_a\left(\boldsymbol{\alpha}_a)\right)\\
    &\downarrow  \\
    \left\{ U_{i}^{(j_i)} \left( \boldsymbol{\theta}_i^{(j_i)}(\boldsymbol{\alpha}_d)\right) \right\}_{j_i\in(1\ldots 4)} &\otimes \left\{ U_{s}^{(j_s)}\left( \boldsymbol{\theta}_s^{(j_s)}(\boldsymbol{\alpha}_a)\right) \right\}_{j_s\in(1\ldots 4)} \,.
\end{split}
\label{eq:vec_pro_comp_var_app}
\end{equation}
Below we list the relations between $\boldsymbol{\alpha}_{d/a}$ and $\boldsymbol{\theta}_{i/s}^{(j)}$ for all $j \in(1\ldots 4)$:
\begin{itemize}
\item $U_{i}^{(1)}$
\begin{equation}
\begin{split}
&  \theta_{i,1}^{(1)} \to \pi/2 \,, \\
&  \theta_{i,2}^{(1)} \to 0 \,, \\
&  \theta_{i,3}^{(1)} \to \pi/2 \,,
\end{split}
\end{equation}
\item $U_{i}^{(2)}$
\begin{equation}
\begin{split}
&  \theta_{i,1}^{(2)} \to 
    \mbox{asin}\left(\cos\alpha_1 \sin\alpha_2 \cos\alpha_3 - 
      \sin\alpha_1 \sin\alpha_3 \right) \,, \\
&  \theta_{i,2}^{(2)} \to \pi/ 2\,, \\
&  \theta_{i,3}^{(2)} \to 
    \mbox{acos}\left( \frac{\cos\alpha_1 \cos\alpha_2}{\sqrt{
     1 - \left(\cos\alpha_1 \sin\alpha_2 \cos\alpha_3 - 
        \sin\alpha_1 \sin\alpha_3\right)^2}} \right) \,,
 \end{split}
\end{equation}
\item $U_{i}^{(3)}$
\begin{equation}
\begin{split}
&  \theta_{i,1}^{(3)} \to 
    \mbox{asin}\left(\sin\alpha_1 \sin\alpha_2 \cos\alpha_3 + 
      \cos\alpha_1 \sin\alpha_3\right)\,, \\ 
&  \theta_{i,2}^{(3)} \to \pi/2 \,, \\
&  \theta_{i,3}^{(3)} \to 
    \mbox{acos}\left( \frac{\sin\alpha_1 \cos\alpha_2}{\sqrt{
     1 - \left(\sin\alpha_1 \sin\alpha_2 \cos\alpha_3 + 
        \cos\alpha_1 \sin\alpha_3\right)^2}}\right) \,,
 \end{split}
\end{equation}
\item $U_{i}^{(4)}$
\begin{equation}
\begin{split}
&  \theta_{i,1}^{(4)} \to 
    \mbox{asin} \left( \cos\alpha_2 \cos\alpha_3\right) \,, \\
&  \theta_{i,2}^{(4)} \to \pi/ 2 \,, \\
&  \theta_{i,3}^{(4)} \to 
    \mbox{acos}\left( \frac{\sin\alpha_2}{\sqrt{
     1 - \left(\cos\alpha_2 \cos\alpha_3\right)^2}}\right) \,,
 \end{split}
\end{equation}
\item $U_{s}^{(1)}$
\begin{equation}
\begin{split}
&  \theta_{s,1}^{(1)} \to 0 \,, \\
&  \theta_{s,2}^{(1)} \to \pi/2 \,, \\
&  \theta_{s,3}^{(1)} \to 0 \,,
\end{split}
\end{equation}
\item $U_{s}^{(2)}$
\begin{equation}
\begin{split}
&  \theta_{s,1}^{(2)} \to \mbox{acos}\left(\cos\alpha_5 \cos\alpha_6 \right) \,, \\
&  \theta_{s,2}^{(2)} \to 
    \mbox{acos} \left( \frac{\sin\alpha_5}{\sqrt{
     1 - \left(\cos\alpha_5 \cos\alpha_6\right)^2}}\right) \,, \\
&  \theta_{s,3}^{(2)} \to \pi/2 \,,
\end{split}
\end{equation}
\item $U_{s}^{(3)}$
\begin{equation}
\begin{split}
&  \theta_{s,1}^{(3)} \to 
    \mbox{acos}\left( \sin\alpha_4 \sin\alpha_5 \cos\alpha_6 + 
      \cos\alpha_4 \sin\alpha_6 \right) \,, \\
&  \theta_{s,2}^{(3)} \to 
    \mbox{acos}\left( \frac{\sin\alpha_4 \cos\alpha_5 }{\sqrt{
     1 - \left(\sin\alpha_4 \sin\alpha_5 \cos\alpha_6 + 
        \cos\alpha_4 \sin\alpha_6\right)^2 }}\right) \,, \\
&  \theta_{s,3}^{(3)} \to \pi/2 \,,
\end{split}
\end{equation}
\item $U_{s}^{(4)}$
\begin{equation}
\begin{split}
&  \theta_{s,1}^{(4)} \to 
  \mbox{acos}\left( \cos\alpha_4 \sin\alpha_5 \cos\alpha_6 - 
    \sin\alpha_4 \sin\alpha_6 \right) \,, \\
&  \theta_{s,2}^{(4)} \to 
  \mbox{acos}\left(\frac{\cos\alpha_4 \cos\alpha_5}{\sqrt{
   1 - \left(\cos\alpha_4 \sin\alpha_5 \cos\alpha_6 - 
      \sin\alpha_4 \sin\alpha_6\right)^2 }}\right)\,, \\
&  \theta_{s,3}^{(4)} \to \pi/2 \,,
\end{split}
\end{equation}
\end{itemize}
where $\mbox{acos}$ and $\mbox{asin}$ are inverse functions of $\cos$ and $\sin$, respectively.
We point out that these choices satisfy our requirements and reproduce the correlation matrix in Eq.~\eqref{eq:X_4case_app}. As already stated, this setting is not unique.



\section*{Declaration statements}

\subsection*{Data Availability}
Data underlying the results presented in this paper are not publicly available, but may be obtained from the authors upon reasonable request.
 
\subsection*{Code Availability}
Not applicable

\subsection*{Acknowledgments}
The work was supported by the Horizon 2020 Framework Programme (899368), Horizon Widera 2023 (101160101) and by the Provincia Autonoma di Trento through the Q@TN joint laboratory.
We kindly thank prof. Mirko Lobino, who lent the two id230 ID-Quantique InGaAs single photon detector modules.

\subsection*{Author Contributions}
A.B. conceived the idea of the quantum algorithm and its photonic implementation. 
A.B. carried out the experiments and the analysis of the results.
A.B. took the lead in writing the manuscript. All authors provided critical feedback and helped shape the research, analysis and manuscript. 

\subsection*{Competing Interests}
Not applicable

\noindent

\printbibliography

\end{document}